\documentclass[journal,draftcls,onecolumn,12pt,twoside]{IEEEtranTCOM}
\IEEEoverridecommandlockouts
\usepackage{acronym}
\usepackage{tikz}
\usetikzlibrary{shapes,snakes}
\usepackage{empheq}
\usepackage{graphicx}
\usepackage{epstopdf}
\usepackage{multirow}
\usepackage{tabu}
\usepackage{array}
\usepackage{url}
\usepackage{amsmath}
\usepackage{amssymb}
\usepackage{amsthm}
\usepackage{amsfonts}
\usepackage{tikz}
\usepackage{bm}
\usepackage{algorithmic}
\usepackage[ruled,vlined,linesnumbered]{algorithm2e}
\usepackage{graphicx}
\usetikzlibrary{arrows}
\usepackage{cite}
\usepackage{caption}
\usepackage{color}
\usepackage{amssymb}
\usepackage{tabulary}
\usepackage{booktabs}

\acrodef{NFV}{Network Function Virtualization}
\acrodef{VNF}{Virtual Network Function}
\acrodef{QBD}{Quasi-Birth-and-Death}
\acrodef{MEC}{Multi-access Edge Computing}
\acrodef{DTMC}{Discrete Time Markov Chain}
\acrodef{QBD}{Quasi-Birth-and-Death}
\acrodef{SNC}{Stochastic Network Caclulus}
\acrodef{SDN}{Software Defined Network}
\acrodef{VM}{Virtual Machine}

\usepackage[justification=centering]{caption} 
\DeclareCaptionLabelFormat{algnonumber}{algorithm}
\tikzstyle{int}=[draw, fill=white!20, minimum size=2em]
\tikzstyle{init} = [pin edge={to-,thin,black}]

\newtheorem{definition}{Definition}

\newcommand\blfootnote[1]{%
	\begingroup
	\renewcommand\thefootnote{}\footnote{#1}%
	\addtocounter{footnote}{-1}%
	\endgroup
}

\newif\iftodo   
\todotrue
\newif\iftodoshort  
\todoshortfalse
\newcommand{\ve}[1]{\boldsymbol{\mathbf{#1}}}

\newcommand{\set}[1]{\mathcal{#1}}

\usepackage{subfig}

\newcommand{\etal}{\textit{et al.}}
\newcommand{\RNum}[1]{\lowercase\expandafter{\romannumeral #1\relax}}

\newtheorem*{remark}{Remark}

\begin{document}
	\title{
	An End-to-End Performance Analysis for Service Chaining in a Virtualized Network 
}
\author{
	\IEEEauthorblockN{Emmanouil Fountoulakis,
		Qi Liao,
		Nikolaos Pappas}
	\vspace{-2.2\baselineskip}}
\maketitle
\begin{abstract}
	\blfootnote{\begin{scriptsize}
			\noindent This work extends the preliminary study in \cite{fountoulakis2019traversing}. This work has been supported by the European Union’s Horizon 2020 research and innovation programme under the Marie Sk\l{}odowska-Curie grant agreement No. 643002. E. Fountoulakis and N. Pappas are with the Department of Science and Technology, Link\"oping University, SE-60174 Norrk\"oping, Sweden (e-mails: \{emmanouil.fountoulakis, nikolaos.pappas\}@liu.se). Q. Liao is with Nokia Bell Labs, 70435 Stuttgart, Germany (e-mail: qi.liao@nokia-bell-labs.com).
	\end{scriptsize}}
Future mobile networks supporting Internet of Things are expected to provide both high throughput and low latency to user-specific services. One way to overcome this challenge is to adopt \ac{NFV} and \ac{MEC}. 
Besides latency constraints, these services may have strict function chaining requirements. In other words, each service has to be processed
by a set of network functions in a specific order. The distribution of network functions over different hosts and more flexible routing caused by service function chaining raise new challenges for end-to-end performance analysis.  
In this paper, as a first step, we analyze an end-to-end communications system that consists of both MEC servers and a server at the core network hosting different types of virtual network functions. We develop a queueing model for the performance analysis of the system consisting of both processing and transmission flows. To approximate the behavior of the system, we decompose the system into subsystems and analyze them independently. By doing so we are able to provide approximate analytical  expressions of the performance metrics such as system drop rate, end-to-end delay, and system throughput. Then, we show how to apply the similar method to an extended larger system and derive a stochastic model for systems with arbitrary number of servers at the edge.
Simulation results show that our approximation model is accurate for the considered systems. We see in Section VI that the simulation and analytical results coincide. By evaluating the system under different scenarios, we provide insights for the decision making on traffic flow control and its impact on critical performance metrics.
\end{abstract}	

\section{Introduction}
The increasing demand of different kinds of network services and the requirements of 5G networks for low capital expenditure raise the need for the improvement of today's networks in terms of both flexibility and scalability. In conventional communications networks, network functions (e.g., firewalls, transcoders, load balancers, etc.) are performed as dedicated hardware middleboxes. Although dedicated hardware can provide high performance, it causes high capital expenditure, low flexibility and scalability, and dependence on particular application. In future communications networks, these limitations will be addressed by Network Function Virtualization (NFV)  \cite{NFV_StateOfTheArt, bonfim2018integrated, li2015software}. The idea of NFV is to decouple the network functions from dedicated hardware equipment. More specifically, general purpose servers can host one or more types of network functions. The network can be flexible and it can deploy proper network functions according to the demands of various traffic types while reducing the capital expenditure. Furthermore, mission-critical mobile applications, e.g., augmented reality, connected vehicles, eHealth, will provide services that require ultra-low latency \cite{MECKeyTech}, \cite{taleb2017multi}. To satisfy such requirements, Multi-access Edge Computing (MEC) has been proposed as a key solution \cite{MECKeyTech}. The idea of MEC is to locate more computation resources closer to the users, e.g., at the base stations.
\ac{NFV} together with MEC are considered key technologies for 5G wireless sytems. However, computation capabilities and  available resources of MEC servers are still limited compared to the high-end servers in the cloud. Therefore, it is interesting to further investigate the cooperation between the edge and the core, and the cooperation among MEC servers.
\subsection{Related work}
Recently, the study of the performance of networks in \ac{VNF}/\ac{SDN} environment has  attracted a lot of attention, \cite{YeE2DDelay, duan2018modeling, miao2019stochastic,  lombardo2014analytical,fahmin2017performance,jarschel2011modeling,goto2016queueing,xiong2016performance}.
Ye \etal \cite{YeE2DDelay} analyze the end-to-end delay for embedded VNF chains. They consider two types of services that traverse different \ac{VNF} chains and provide the delay analysis for each different chain.
Miao \etal \cite{miao2019stochastic}  provide an analytical model based on \ac{SNC} to provide upper and lower delay bounds  of a \ac{VNF} chain. In their analysis, they consider both the case of bursty and non-bursty traffic. Along similar lines, Duan \cite{duan2018modeling} analyzes an end-to-end delay performance of service function chaining for particular services and given resources.
Authors in \cite{lombardo2014analytical,fahmin2017performance,jarschel2011modeling, goto2016queueing,xiong2016performance},  apply tools from queueing theory to evaluate the performance of systems in \ac{SDN} environment. In particular, Jarschel  \etal \cite{jarschel2011modeling} study the OpenFlow architecture, where the switch is modeled as an M/M/1 queue and the controller as a feedback system of the delay-loss type M/M/1/S queue. Similarly Goto \etal \cite{goto2016queueing} analyze a simple OpenFlow-based switch in SDN environment, however, they distinguish traffic from the controller and exogenous traffic. 
Furthermore, a reasonable amount of works consider the modeling of connected \acs{VNF} as a sequence of queues where the goal is to guarantee the stability of the system and some particular network or service requirements. A well known and widely used mathematical tool for stabilizing dynamic systems is the Lyapunov optimization theory.
The authors in \cite{feng2018optimal, barcelo2015cloud, feng2016dynamic, feng2018optimalwireless, chen2019automated,gu2019fairness} develop dynamic algorithms by applying Lyapunov optimization theory in order to guarantee system stability and fulfill additional service requirements. In particular, in \cite{feng2018optimal, feng2018optimalwireless} flows traverse \ac{VNF} chains and each node decides the resource allocation to each \ac{VNF} and routes the flow to the next node. Gu \etal \cite{gu2019fairness} develop a dynamic distributed algorithm that controls the flow and rate at each node. Their objective is to achieve fairness between the services and maximize the network utility while providing system stability. Chen \etal \cite{chen2019automated} consider the problem where \ac{VNF} are installed in \acp{VM}. Each  \ac{VM} can be located in the same or different data center. In this work, each \ac{VM} decides which functions to install or uninstall and which services to serve. A dynamic algorithm is developed  that works in a distributed manner and takes online decisions.
Considering a more static and known environment, researchers investigate the \ac{VNF} placement and resource allocation problems \cite{NearOptPlac, JointOptJournal,qu2016delay, xu2019NFVenabled, gouareb2018virtual}. In these works, the authors formulate the VNF placement problem as Mixed Integer Linear Problem (MILP) under the assumption of known traffic demand.  In addition, approximation algorithms have been developed in order to provide a solution to the VNF placement problem, see for example  \cite{feng2017approximation, lange2017multi}. 

Besides the \ac{VNF}/\ac{SDN} technologies, \ac{MEC} technology promises a significant improvement of the networks, especially in terms of latency reduction. Task offloading from the mobile devices to \acp{MEC} has attracted considerable attention. For example, the authors in \cite{liu2016delay, mao2017stochastic, HanPowerOptimalScheduling, lyu2018energy}  address the trade-off between the power consumption and the task processing delay. The authors in \cite{xiangnetworkslicing}, propose an optimization model to perform joint slicing of mobile network and edge computation resources.
As VNF and MEC are two strongly connected technologies, the research evolution of the one affects the evolution of the other. To the best of our knowledge, there are no works providing end-to-end network performance analysis for a network with MEC deployed and operated within the NFV environment.
\subsection{Contributions}
In this paper, we investigate a \ac{VNF}-facilitated end-to-end network where the cooperation between the edge and the core network, and the cooperation among \acp{MEC} are considered. As a first step, we model and analyze a simple end-to-end communications system which consists of two MEC servers at the edge network and one at the core network hosting different types of VNFs. We develop a methodology by applying tools from queueing theory in order to derive approximate analytical expressions of the performance metrics of interest such as end-to-end delay, drop rate, and throughput. The methodology is based on the decomposion of the system into smaller subsystems easier to analyze. The information from analysis of each subsystem can be used in order to study the performance of the entire system. Since, in reality, there are no buffers with infinite capacity, we consider finite size buffers in the servers of our system model. However, the capabilities of the server at the core are considered to be much higher than those of the MEC servers. In order to observe the behavior of the system asymptotically, we also investigate, as a subcase, the scenario where the buffers at  core server is infinite. Based on the methodology derived for the simple model, we extend the analysis to a more complex system with a larger number of MEC servers. As shown in the simulation results, the proposed stochastic model is accurate and indicates a robust behavior even for larger number of servers. Furthermore, we provide results that show the trade-off between the throughput, drop rate, and end-to-end delay. In this work we provide approximate analytical expressions for the  performance metrics of our interest in order to provide  insights of how to design a system to satisfy particular network requirements. 

\section{System Model}
The aim of this work is to study the performance of an end-to-end communications system that consists of different types of \acp{VNF}. We first provide the analysis for a simple end-to-end communications system and then, we provide the analysis for more general topologies. 

\begin{figure}[t!]
	\centering
	\includegraphics[scale=0.4]{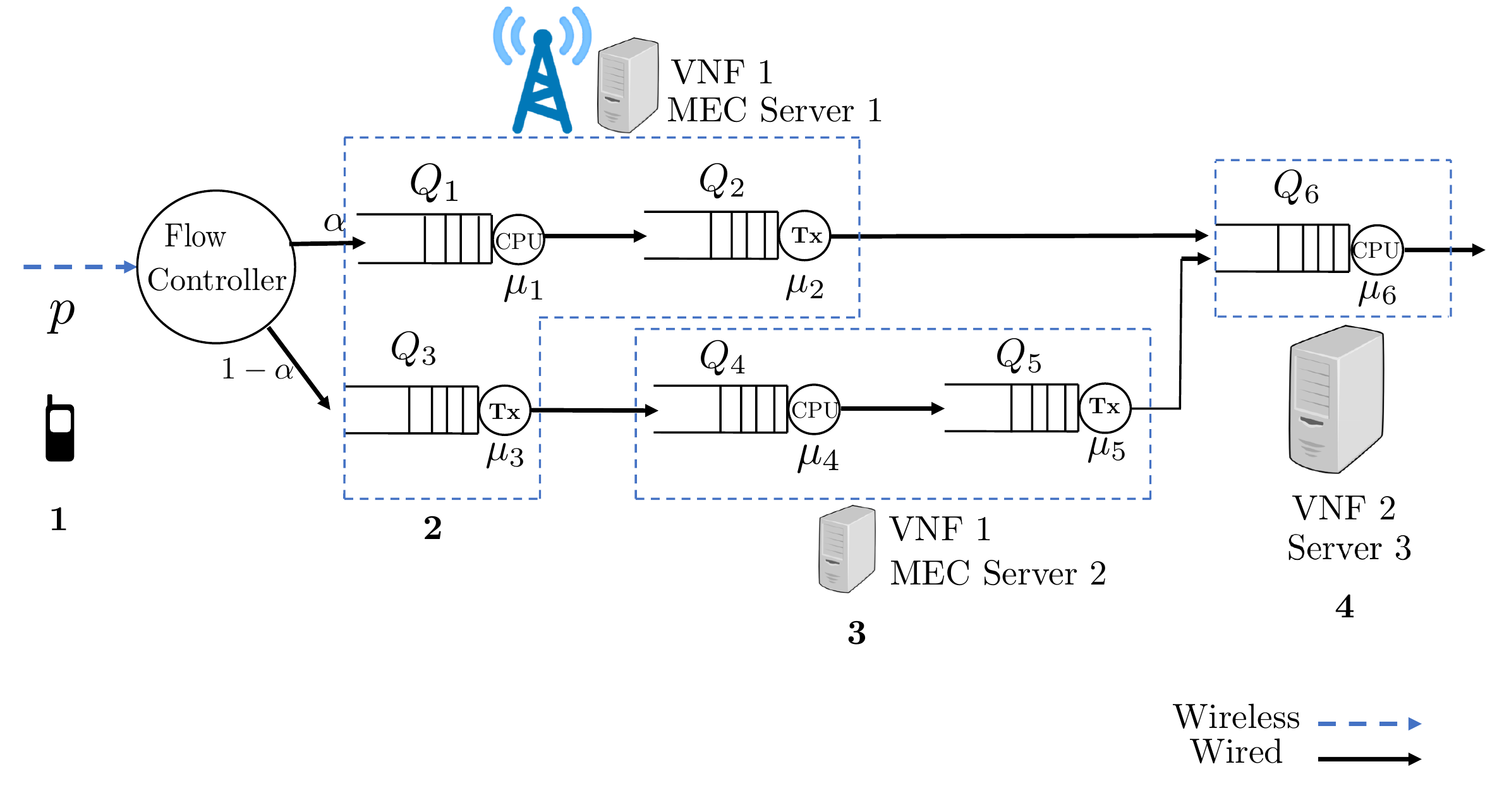} 
	\caption{System model: Blue dashed lines group the queues located in the same server.}
	\label{Fig: System}
\end{figure}

We consider an end-to-end communications system consisting of a mobile device, two \ac{MEC} servers, and one server located in the core network as depicted in Fig. \ref{Fig: System}.
A task traverses a service chain of two consecutive VNFs: VNF $1$ and VNF $2$. In the chosen system, a \ac{MEC} server, called Server $1$, is co-located with the base station and hosts one copy of VNF $1$ as the primary MEC server. A secondary \ac{MEC} server, called Server $2$, is located nearby and also hosts a copy of VNF $1$. Server $2$, for example, may be located at a peer edge host with spare capacity or at a central office location within a metropolitan area network. In addition, Server $3$ in the core network hosts VNF $2$ and has more advanced computation capabilities than Servers $1$ and $2$. The analysis of such system can be used to study the deployment of AR-related applications. For example, VNF function Domain called Name Service (DMS) can be installed in MEC to map user requests to their corresponding nodes providing contents, while the compute-intensive VNFs such as AR or 3D gaming can be deployed in core networks. 

We assume a slotted time system. At each time slot, the device transmits a task in form of a packet to a base station over a wireless channel. Because of the presence of fading in the wireless channel, transmissions may face errors. Thus, we assume that a task is successfully transmitted to the base station with a probability $p$ that captures fading, attenuation, noise, etc. The device attempts for a new task transmission only if the previous task is successfully received at the base station.
The received  tasks need to be distributed between the queue for local processing and the queue for transmission to the secondary MEC server.

Thus, there are two possible routes to pass through the service chain.
A flow controller at the base station decides randomly the routing for each task. With probability $\alpha$ the task is processed by Server $1$ first to be processed by VNF $1$, and then forwarded to Server $3$ to be processed by VNF $2$. With probability $1-\alpha$ the task is forwarded to Server $2$, to be processed by VNF $1$, and then forwarded to Server $3$ for being processed by VNF $2$.   

Each task that arrives at a server first waits in a queue for being processed by a \ac{VNF}. Then, after the processing, it is stored in the transmission queue, waiting to be forwarded and processed by the next \ac{VNF}. 
Let $Q_{i}$ denote the $i$-th queue, where $i \in \mathcal{K}$, and $\mathcal{K}$ is the set of the queues in the system.
Note that the queues follow an early departure-late arrival model: at the beginning of the slot the departure takes place and a new arrival can enter the queue at the end of the slot. 
The queues for task transmission are $Q_{2}$, $Q_{3}$, and $Q_{5}$, and the queues for task processing are $Q_{1}$, $Q_{4}$, $Q_{6}$. 
The arrival rates for queues $Q_1$ and $Q_3$ are $p\alpha$ and $p(1-\alpha)$, respectively. We denote by $\mu_{i}$, $i \in\set{K}$, the service rates of the queues. We assume that the service times are geometrically distributed. Furthermore, given that $Q_1$, $Q_3$, and $Q_4$ are non empty, the arrival rates of $Q_2$, $Q_4$, and $Q_5$ are equivalent to the service rates of $Q_1$, $Q_3$, and $Q_4$ (i.e., $\mu_1$, $\mu_3$, and $\mu_4$) respectively. 

Furthermore, the queues are assumed to have finite buffer. Let $M_{i}$ denote the buffer size of each queue $i \in \mathcal{K}$. If a queue is full and no task departs  at the same time that a new one arrives, the new task is dropped and removed from the system. Besides the case presented in Fig. 1, we will consider the case of a scaled-up system in Section V. Note that the described system model can be considered as an isolated network slice under the network slicing and \ac{MEC} technologies.  

\section{Performance Analysis}\label{Sec: PerAnalysis}
In this section, we perform the modeling and the performance analysis that allow us to derive the critical  performance metrics. We model the considered queueing system utilizing \ac{DTMC}. Modeling the whole system as one Markov chain can drive in a quite complicated system difficult to be analyzed in terms of closed-form expressions. Thus, in order to simplify the analysis, we decompose the system
into different subsystems. Since the computation capabilities of the server in the core are much higher than those of the \ac{MEC} servers, we consider, in the analysis, the core server as one independent system. In order to develop a model that couples MEC Server $1$ and MEC Server $2$, we consider as one independent system the subsystem which consists of $Q_{3}$ and $Q_{4}$. $Q_{5}$ is considered as one independent system. We choose to decompose the system as described above in order to study the decomposition strategies where derive: \RNum{1}) a model for the interacting queues within one server, \RNum{2}) a model for two queues of different servers that are interacting, and \RNum{3}) an individual model for each queue within the same server.
To summarize, we consider the following four subsystems: 1) $Q_1$ and $Q_2$, 2) $Q_3$ and $Q_4$ 3) $Q_5$, and 4) $Q_6$. The performance metrics for the whole system are approximated with the analytical expressions derived from the subsystems. The accuracy of the approximation is validated through simulations in Section \ref{Sec:Sim}.
\subsection{Subsystems 1 and 2: Two queues in tandem}
\begin{figure}[t!]
	\centering
	\includegraphics[scale=0.4]{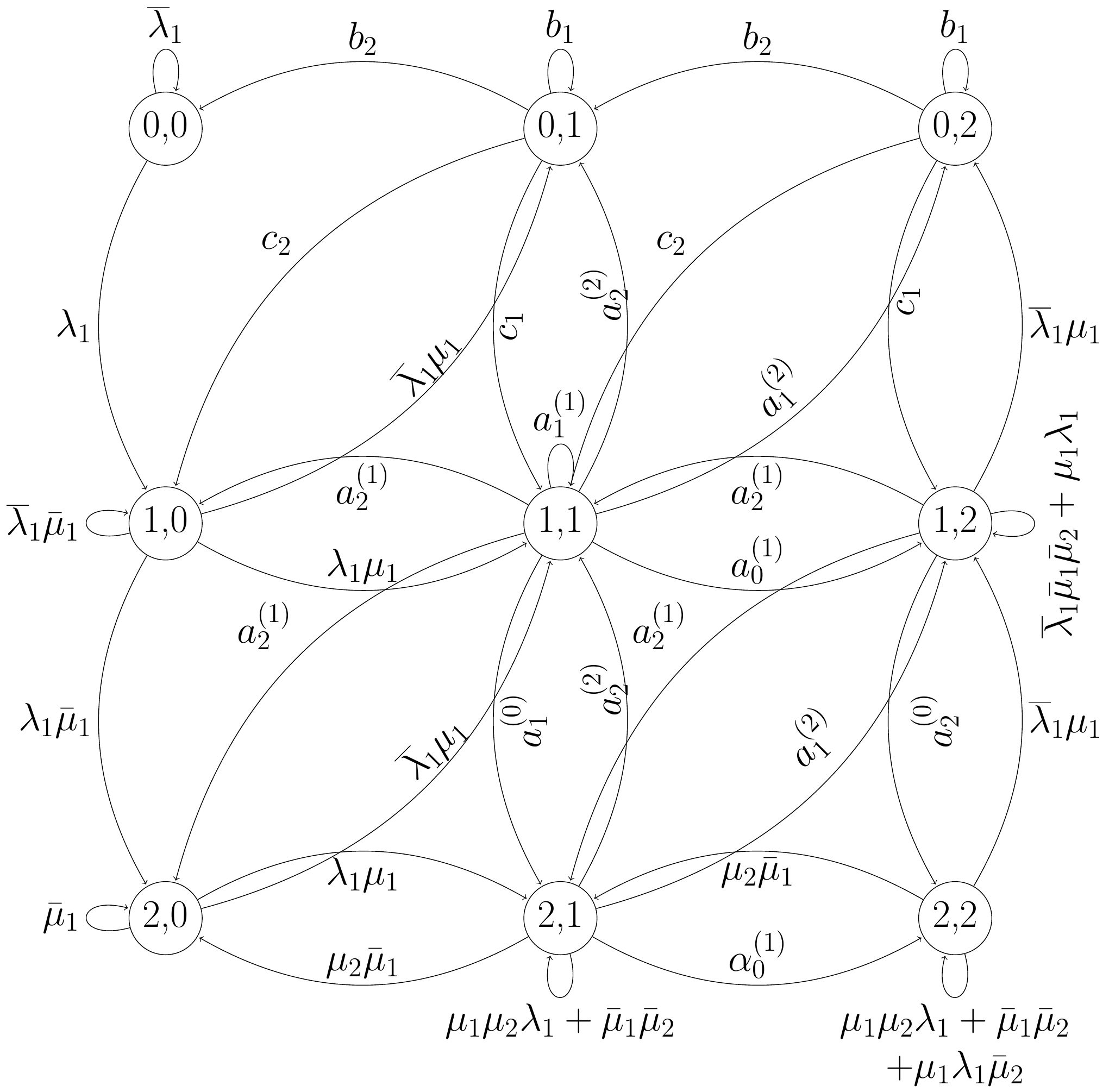}
	\caption{Two dimensional Markov chain for the subsystem consisting of $Q_1$ and $Q_2$.
	}\label{Fig: 2dMarkovChain} 
\end{figure}

The two queues in tandem, $Q_{1}$ and $Q_{2}$, are considered as one subsystem. The arrival rate for $Q_{1}$ is: $\lambda_{1}=p\alpha$.
The Markov chain $\left\{(X_{n}\text{, } Y_{n})\right\}$ is described by $$P_{i,j:u,k} = \Pr\left\{X_{n+1}=i\text{, } Y_{n+1} = j\text{ }|\text{ } X_{n}=u,Y_{n}=k\right\}\text{,}$$ where $X_n$ and $Y_n$ denote the states (in terms of queue length) of $Q_1$ and $Q_2$ at the $n$-th time slot, respectively; $i$ and $j$ are referred to as the level $i$ and phase $j$, respectively. 
In order to facilitate the presentation, we first analyze a simple example with buffer size $M_{1}=M_{2}=2$. The Markov chain of this example is shown in Fig. \ref{Fig: 2dMarkovChain}. However, the analysis presented below is quite general and independent of the specific buffer size. The Markov chain is a \ac{QBD} \ac{DTMC} \cite{AppliedDiscrete}. Note that the \ac{QBD} only goes a maximum level up or down, the transition matrix has a block partitioned form:
\begin{align}\label{Eq: TranP1}
\mathbf{P}_{1}=
\left[\begin{array}{ccccc}
\mathbf{B} & \mathbf{C} & & &  \\
\mathbf{E} & \mathbf{A}_{1} & \mathbf{A}_{0} & &  \\
& \mathbf{A}_{2} & \mathbf{A}_{1} & \mathbf{A}_{0}  & \\
& & \ddots & \ddots &    \\
&&& \mathbf{A}_{2}  & \mathbf{A}_{0}+\mathbf{A}_{1} 
\end{array}\right]\text{.}
\end{align}	
For simplicity, given a probability of an event, denoted by $p$, we denote the probability of its complementary event by $\bar{p}\triangleq 1-p$. 
The block matrices of $\mathbf{P}_{1}$ are shown below

\begin{align}
\mathbf{B}=
\left[\begin{array}{ccc}
\overline{\lambda}_{1} &  0  &   0 \\
b_2 & b_1 &  0   \\
0  &  b_2      &   b_1
\end{array}\right]\text{, }
\mathbf{C}=
\left[\begin{array}{ccc}
\lambda_{1}  &  0   &  0 \\
c_2  &  c_1  &  0   \\\nonumber
0  &  c_2  & c_1  
\end{array}\right]\text{, }
\mathbf{E}=
\left[\begin{array}{ccc}
0  &  \overline{\lambda}_{1} \mu_{1}      &                 0                   \\
0  &  \overline{\lambda}_{1} \mu_{1}\mu_{2}  &  \overline{\lambda}_{1} \mu_{1} \bar{\mu}_{2}\\ \nonumber
0  &  0  &  \overline{\lambda}_{1} \mu_{1}  
\end{array}
\right]\text{,}
\end{align}

\begin{align}
\mathbf{A}_{0}=
\left[\begin{array}{ccc}
\lambda_{1} \bar{\mu}_{1}   &  0  &  0\\
a_{2}^{(0)} &  a_{1}^{(0)} & 0 \\ \nonumber
0  &  a_{2}^{(0)}   &  a_{1}^{(0)}
\end{array}\right]\text{, }
\mathbf{A}_{1}=
\left[\begin{array}{ccc}
\overline{\lambda}_{1}  &  \lambda_{1} \mu_{1}  & 0\\ 
a_{2}^{(1)}              &  a_{1}^{(1)}           & a_{0}^{(1)} \\ 
0                            &  a_{2}^{(1)}          & a_{1}^{(1)}  
\end{array}\right]\text{, }
\mathbf{A}_{2}=
\left[\begin{array}{ccc}
0  &  \bar{\lambda}_{1}\mu_{1} &  0 \\
0  &  a_{2}^{(2)}  &  a_{1}^{(2)} \\ \nonumber
0  &  0  &  a_{2}^{(2)} 
\end{array}\right]\text{.} 
\end{align}

We construct the Markov chain of this particular example. Let us define $b_1=\overline{\lambda}_{1}\bar{\mu}_{2}$, $b_2=\overline{\lambda}_{1}\mu_{2}$, $c_1=\lambda_{1}\bar{\mu}_{2}$, $c_{2}=\lambda_{1} \mu_{2}$, $a_{1}^{(0)}=\lambda_{1}\bar{\mu}_{1}\bar{\mu}_{2}$, $a_{2}^{(0)}=\lambda_{1}\bar{\mu}_{1}\mu_{2}$, $a_{0}^{(1)}=\lambda_{1} \mu_{1}\bar{\mu}_{2}$, 
$a_{1}^{(1)}=\lambda_{1} \mu_{1}\mu_{2} + \overline{\lambda}_{1} \bar{\mu}_{1}\bar{\mu}_{2}$, $a_{2}^{(1)}=\overline{\lambda}_{1}\bar{\mu}_{1}\mu_{2}$, and
$a_{1}^{(2)}=\overline{\lambda}_{1}\mu_{1}\bar{\mu}_{2}$, $a_{2}^{(2)}=\overline{\lambda}_{1} \mu_{1} \mu_{2}$. Then, we construct the transition matrix of the Markov chain and observe a particular structure of the matrix because of the properties of a QBD DTMC.
By utilizing the properties of a QBD DTMC, we can analyze such systems with arbitrary buffer size in the following steps.
First we define the following  matrices \cite{AppliedDiscrete}
\begin{align}\nonumber
\mathbf{P}_{1}^{(1)}=
\left[\begin{array}{cccc}
1 & 0 & &\\
\mu_{2} & \bar{\mu}_{2}  & &\\
& \ddots & \ddots &       \\ 
& &\mu_{2} &   \bar{\mu}_{2}  
\end{array}    
\right]\text{, }
\mathbf{P}_{1}^{(2)}=
\left[\begin{array}{ccccccc}
0 & 1 & 0 & &\\
0 & \mu_{2} & \bar{\mu}_{2}  & &\\
0 &  0 & \mu_{2}  &  \bar{\mu}_{2}  &\\
& & & \ddots  &  \ddots   &  \\
& & &         &    0       &      1  
\end{array}
\right]\text{.}
\end{align}
Then, the block matrices of the transition matrix are calculated as  \cite{AppliedDiscrete}
$$\mathbf{B}= \bar{\lambda}_{1}\mathbf{P}_{1}^{(1)}\text{, } \mathbf{C}=\lambda_{1}\mathbf{P}_{1}^{(1)}\text{, } \mathbf{E}=\bar{\lambda}_{1}\mu_{1}\mathbf{P}_{1}^{(2)}\text{,}$$
$$\mathbf{A}_{0} = \lambda_{1}\bar{\mu}_{1}\mathbf{P}_{1}^{(1)}\text{, } \mathbf{A}_{1}=\bar{\lambda}_{1}\bar{\mu}_{1}\mathbf{P}_{1}^{(1)}+\lambda_{1}\mu_{1}\mathbf{P}_{1}^{(2)}\text{, } \mathbf{A}_{2}=\bar{\lambda}_{1}\mu_{1}\mathbf{P}_{1}^{(2)}\text{.}$$ 
\noindent Following the steps described above, we can construct the transition matrix of Subsystem $1$ for arbitrary finite buffer sizes.
Our goal is to derive the steady state distribution of the Markov chain defined above. We can apply direct methods in order to find the steady state distribution \cite[Chapter 4]{AppliedDiscrete}. Note that there are several efficient algorithms that can be used for this purpose, e.g., logarithmic reduction method \cite{AppliedDiscrete}. 
We denote the steady state distribution of Subsystem $1$ by a row vector defined as $\boldsymbol{\pi}^{(1)} = \left[\pi_{0,0}^{(1)},\pi_{0,1}^{(1)},\ldots,\pi_{0,M_{2}}^{(1)},\pi_{1,0}^{(1)},\ldots,\pi_{M1,M2}^{(1)}\right]\text{.}$ We find $\boldsymbol{\pi}^{(1)}$ by solving the following linear system of equations
$
\boldsymbol{\pi}^{(1)} \mathbf{P}_{1} =\boldsymbol{\pi}^{(1)}\text{, }  
\boldsymbol{\pi}^{(1)} \boldsymbol{1} = 1\text{,}
$
where $\boldsymbol{1}$ denotes the column vector of ones. Hereafter we use $\ve{\pi}^{(n)}$ to denote the steady state distribution vector of the $n$-th subsystem for $n = 1, 2, 3, 4$. 

Furthermore, the arrival rate of $Q_{2}$ depends on the service rate of $Q_{1}$. However, the arrival rate of $Q_{2}$ is equal to $\mu_{1}$ if and only if $Q_{1}$ is non-empty. Therefore, we define the arrival rate of $Q_{2}$ as
$
\lambda_{2} = \text{Pr}\left\{Q_{1}>0\right\} \mu_{1} = \left(\sum_{j=0}^{M_2} \sum_{i=1}^{M_{1}} \pi_{i,j}^{(1)}\right) \mu_{1}\text{.}
$ Similarly, we can construct the transition matrix $\mathbf{P}_{2}$ and the steady state distribution $\boldsymbol{\pi}^{(2)}$ for the second subsystem consisting of $Q_{3}$ and $Q_{4}$. The arrival rates of $Q_{3}$ and $Q_{4}$ are
\begin{flalign*}
\lambda_{3}  = p (1-\alpha)  \text{ and }
\lambda_{4}   = \text{Pr} \left\{Q_{3}>0\right\} \mu_{3} = \left( \sum\limits_{j=0}^{M_{4}} \sum\limits_{i=1}^{M_{3}} \pi_{i,j}^{(2)} \right) \mu_{3}\text{, respectively.}
\end{flalign*}

\subsection{Subsystem 3:  $Q_{5}$} 
\begin{figure}
	\centering
	\includegraphics[scale=0.6]{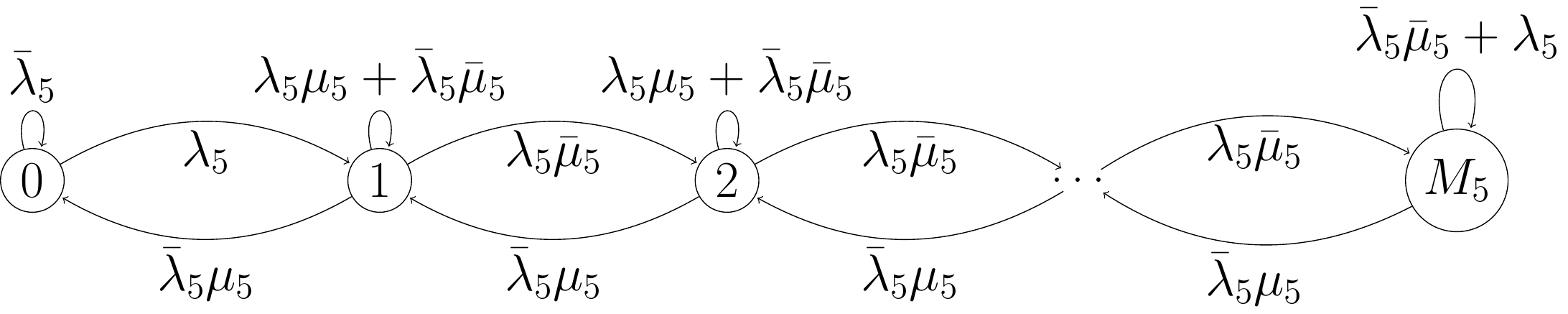}
	\caption{Markov chain for  $Q_5$.}
	\label{Fig: MarkovQBDone}
\end{figure}

We consider $Q_5$ as an independent subsystem. We denote by $M_{5}$ the buffer size of the queue. We first define the arrival rate of $Q_{5}$ as
$\lambda_{5} = \text{Pr}\left\{Q_4>0\right\} \mu_{4} = \left(\sum\limits_{i=0}^{M_{3}} \sum\limits_{j=1}^{M_{4}} \pi_{i,j}^{(2)}\right) \mu_{4}\text{.}$
The Markov chain of this system is shown in Fig. \ref{Fig: MarkovQBDone}. 
The transition matrix of this Markov chain is described below
\begin{align}\nonumber
& \mathbf{P}_{3} = \left[\begin{array}{cccccc}
\bar{\lambda}_{5}           & \lambda_{5}                                                       &  &  \\
\bar{\lambda}_{5}\mu_{5}  & \lambda_{5}\mu_{5}+\bar{\lambda}_{5}\bar{\mu}_{5} & \lambda_{5}\bar{\mu}_{5}  & \\
& \bar{\lambda}_{5}\mu_{5} & \lambda_{5}\mu_{5}+\bar{\lambda}_{5}\bar{\mu}_{5} & \lambda_{5}\bar{\mu}_{5}  \\
& \ddots & \ddots & \ddots  &     \\
&            &  \bar{\lambda}_{5}\mu_{5} &  \bar{\lambda}_{5}\bar{\mu}_{5} + \lambda_{5}     
\end{array}
\right]\text{.}
\end{align}
We denote the steady state distribution of Subsystem 3 by $\boldsymbol{\pi}^{(3)}= \left[\pi_{0}^{(3)}\text{, }\pi_{1}^{(3)}\text{,}\ldots\text{, } \pi_{M_5}^{(3)}  \right]$.
To derive $\mathbf{\pi}^{(3)}$, we solve the following linear system of equations,
$
\boldsymbol{\pi}^{(3)}\mathbf{P}_{3} = \boldsymbol{\pi}^{(3)}\text{, } 
\boldsymbol{\pi}^{(3)}\mathbf{1} = 1\text{.}
$

\noindent Using balance equation, we obtain 
\begin{align}\nonumber
\pi^{(3)}_{i} = \frac{\lambda_{5}^i\bar{\mu}_{5}^{(i-1)}}{\bar{\lambda}_{5}^{i}\mu_{5}^{i}}\pi_{0}^{(3)}\text{, } \text{for } i=1\text{, }\ldots \text{, } M_{5}\text{ and } \pi^{(3)}_{0} = \left[1+\sum\limits_{i=1}^{M_{5}}\frac{\lambda_{5}^{i}\bar{\mu}_{5}^{i-1}}{\bar{\lambda}_{5}^{i}\mu_{5}^{i}} \right]^{-1}\text{.}
\label{eqn:BalanceEqn_Q5}
\end{align}

\subsection{Subsystem $4$: $Q_6$ with finite buffer size}
\begin{figure}[t!]
	\centering
	\includegraphics[scale=0.6]{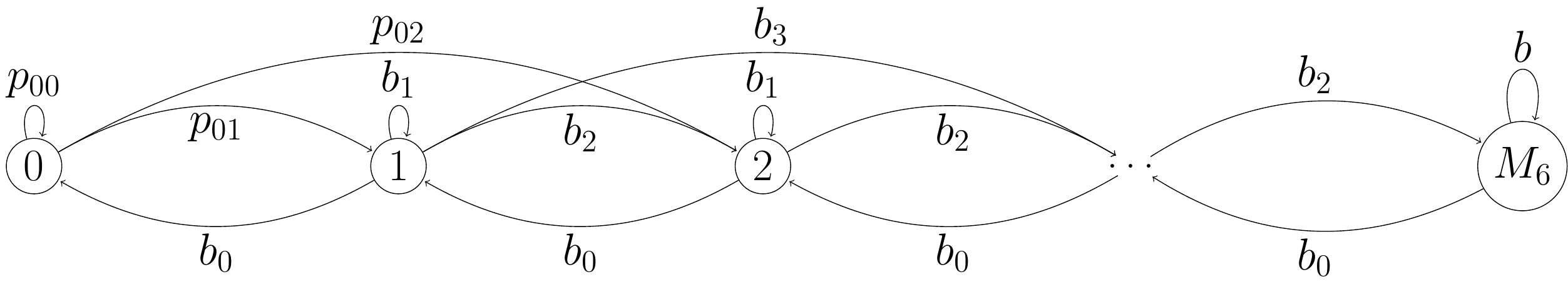}
	\caption{Markov chain for $Q_6$.}\label{Fig: 2BernMarkov}
\end{figure}

The arrival rate for $Q_6$ depends on the service rate of $Q_{2}$ and $Q_{5}$, and the probability the queues to be non-empty. Note that the departures from $Q_{2}$ and $Q_{5}$ can be considered independent stochastic processes. The arrival rates for $Q_{6}$ that occur due to $Q_{2}$ and $Q_{5}$ are 
$\lambda_{6,2}   = \text{Pr}\left\{Q_{2}>0\right\}\mu_{2}\text{, } \text{and } \lambda_{6,5} = \text{Pr}\left\{Q_{5}>0\right\} \mu_{5} \text{, respectively.}$
The arrival rate of $Q_{6}$ is given by $ \lambda_{6} = \lambda_{6,2} + \lambda_{6,5}\text{.}$
We model the system as a Markov chain as shown in Fig. \ref{Fig: 2BernMarkov}, where
\begin{align}\nonumber 
p_{00} & = \bar{\lambda}_{6,2}\bar{\lambda}_{6,5}\text{, }p_{01}  = \lambda_{6,2}\bar{\lambda}_{6,5} + \lambda_{6,5}\bar{\lambda}_{6,2}\text{, } 	p_{02}  = \lambda_{6,2}\lambda_{6,5}\text{, }\\\nonumber
b_{1} & = \bar{\lambda}_{6,5}\bar{\lambda}_{6,2}\bar{\mu}_{6}+\bar{\lambda}_{6,5}\lambda_{6,2}\mu_{6} + \lambda_{6,5}\bar{\lambda}_{6,2}\mu_{6}\text{, } \\ \nonumber
b_{2}  & =  \lambda_{6,2}\bar{\lambda}_{6,5}\bar{\mu}_{6} + \bar{\lambda}_{6,2}\lambda_{6,5}\bar{\mu}_{6} + \lambda_{6,2}\lambda_{6,5}\mu_{6}\text{,}\\\nonumber
b_{3}  &  = \lambda_{6,2}\lambda_{6,5}\bar{\mu}_{6}\text{, } b_{0}   = \bar{\lambda}_{6,2}\bar{\lambda}_{6,5}\mu_{6}\text{,}\\\nonumber
b & = \bar{\lambda}_{6,2}\lambda_{6,5}\mu_{6} + \lambda_{6,2}\bar{\lambda}_{6,5}\mu_{6} + \bar{\lambda}_{6,2}\bar{\lambda}_{6,5}\bar{\mu}_{6}\text{.}
\end{align}

The transition matrix that describes the Markov chain above is shown below
\begin{align}
\mathbf{P}_{4} = 
\left[\begin{array}{ccccccc}
p_{00}    &      p_{01}    &       p_{01}     &                          \\
b_{0}          &      b_{1}          &        b_{2}          &         b_{3}      &    \\
&      b_{0}         &         b_{1}          &      b_{2}         & b_{3}      \\
&                       &       \ddots         &    \ddots       & \ddots & \ddots \\ 
&       &                        &         b_{0}	   	    &          b_{1}   &    b_{2}    & b_{3} \\
&                       &          &             &    b_{0}    & b_{1}   &   b_{2} + b_{3} \\
&                       &          &            &                 &    b_{0}       &    b_{1} + b_{2} + b_{3}
\end{array} 
\right]\text{.}
\end{align}
In order to construct the transition matrix $\mathbf{P}_{6}$, we need only to calculate the probabilities described above. We denote the steady state distribution of Subsystem $4$ by $\boldsymbol{\pi}^{(4)}= \left[\pi_{0}^{(4)}\text{, }\pi_{1}^{(4)}\text{, }\ldots\right]$.
To derive $\bm{\pi}^{(4)}$, we solve the following linear system of equations
\begin{align}
\bm{\pi}^{(4)} \mathbf{P}_{4} = \bm{\pi}^{(4)}\text{, } \bm{\pi}^{(4)}\bm{1}=1\text{.}
\label{eq: LinSys6}
\end{align}
We observe from (\ref{eq: LinSys6}) that $\bm{\pi}^{(4)}$ is the eigenvector of the transition matrix for $\lambda=1$. Therefore, we can use eigenvalue decomposition (EVD) in order to find the eigenvectors of the matrix. After applying EVD, the transition matrix can be written as
$
\mathbf{P}^{(4)} = \mathbf{Q}\mathbf{\Lambda}\mathbf{Q}^{-1}\text{,}
$
where $\mathbf{Q} = \left[\mathbf{q}_{1}, \ldots\mathbf{q}_{n}\right]$ are the eigenvectors, and matrix $\mathbf{\Lambda}$ contains the eigenvalues in its diagonal. The  eigenvector that corresponds to the the eigenvalue that has value equal to one, is the steady state distribution. 

\subsection{Subsystem 4: $Q_{6}$ with infinite buffer size}
In this subsection, we study the case where the buffer in the core server has infinite size. Since the capabilities of the server in the core can be much higher than those of the MEC server and therefore, the buffer has a very large capacity, we are interested in studying the stability conditions of such systems. We model the system as one Markov chain. 
The transition matrix that describes the Markov chain  is shown below
\begin{align}
\mathbf{P}_{4} = 
\left[\begin{array}{ccccccccc}
a_{0}       &      b_{0}     &       0          &         0      &  \cdots  \\
a_{1}       &      b_{1}     &     b_{0}      &         0      & \cdots   \\
a_{2}      &      b_{2}     &     b_{1}       &      b_{0}   & \cdots  \\
0           &     b_{3}      &     b_{2}      &      b_{1}    & \cdots \\ 
&       0          &     b_{3}      &      b_{2}    &   b_{1}  & \\
&                 &        \ddots  &   \ddots       &      \ddots  & \ddots
\end{array}
\right]\text{,}
\end{align}
where $a_{0}=p_{00}$, $a_{1}=p_{01}$, $a_{2}=p_{02}$. The transition matrix is a lower Hessenberg matrix. 
The general expression for the equilibrium equations for the states is given by the $i$-th term in the following equation:
$
\pi_{i}^{(4)}=a_{i}\pi_{0}^{(4)}+\sum\limits_{j=1}^{i+1} b_{i-j}\pi_{j}^{(4)}\text{.}
$
For the \ac{DTMC} with infinite state space, we apply $z$-tansform approach to solve the state equations. The $z$-transform for the state transition probabilities $a_{i}$ and $b_{i}$ are
$
A(z) = \sum\limits_{i=0}^{2} a_{i}z^{-i}\text{ and }
B(z) = \sum\limits_{i=0}^{3}b_{i}z^{-i}\text{,}
$ respectively. The $z$-transform for the steady state distribution vector $\boldsymbol{\pi}^{(4)}$ is
$
\Pi(z) = \sum\limits_{i=0}^{\infty}\pi_{i}^{(4)} z^{-i}=\pi_{0}^{(4)}\frac{z^{-1}A(z)-B(z)}{z^{-1}-B(z)}\text{.}
$
The solution for $\pi_{i}^{(4)}$ is given by
$
\pi_{0}^{(4)}  = \frac{1+B'(1)}{1+B'(1)-A'(1)}\text{, }
\pi_{i}^{(4)}   =c_{i}  +  \sum\limits_{j=1}^{m} r_{j}(p_{j})^{(i-1)}\text{, } i>0\text{,}
$
where $r$, $p$, and $c$ are the residues, poles, and directs terms, respectively.
Since $Q_6$ has infinite buffer size, we need to characterize the conditions under which the queue is stable.
Stability is important since it implies finite queueing delay. A definition of queue stability is shown below \cite{szpankowski1994stability}.

\begin{definition}
	Denote by $Q_{i}(t)$ the length of queue $i$ at the beginning of time slot $t$. The queue is said to be stable if  $\lim\limits_{t\rightarrow \infty} \Pr\left\{Q_{i}(t)<x\right\} = F(x)$ and $\lim\limits_{x\rightarrow \infty}F(x)=1$.
\end{definition}

The corollary consequence of the previous definition is the Loynes' theorem \cite{loynes1962stability} that states: if the arrival and service processes of a queue are strictly jointly stationary and the average arrival rate is less than the average service rate, then the queue is stable.
Therefore, $Q_6$ is stable if and only if the following inequality holds: $\lambda_{6}<\mu_{6}\text{.}$

\begin{remark}
	Since the stationary distributions of the presented Markov chains have been calculated, we can also obtain the distributions of the queue sizes. Thus, we can write the probability that a queue size goes beyond a congestion limit. For example, consider $C$ as the congestion limit for $Q_{5}$, the congestion violation probability is calculated as $\Pr \left\{Q_{5} > C \right\} = \sum\limits_{i=C+1}^{M_{5}}\pi^{(3)}_{i}$.
\end{remark}

\section{Key Performance Metrics}\label{Sec: KPM}
In this section, we provide analytical expressions of the  performance metrics of our interests. First, we calculate the system drop rate and average number of tasks in the system that are necessary metrics for analyzing the throughput and delay of the system. We utilize the results of the previous section in order to obtain the corresponding expressions. 
\subsection{Drop rate and average queue length}
The probabilities to have a dropped task at each time slot for $Q_1-Q_5$ are shown respectively in below
\begin{align}\nonumber
P_{D_{1}} & = \lambda_{1} \bar{\mu}_{1}\sum\limits_{j=0}^{M_2} \pi^{(1)}_{M_{1},j}\text{, }P_{D_{2}}  =\lambda_{2}\bar{\mu}_{2}\sum\limits_{i=1}^{M_1}\pi^{(1)}_{i,M_{2}} \text{, }
\end{align}
\begin{align}\nonumber
P_{D_{3}} & = \lambda_{3}\bar{\mu}_{3} \sum\limits_{j=0}^{M_4} \pi^{(2)}_{M_{3},j}\text{, }
P_{D_{4}}  = \lambda_{4}\bar{\mu}_{4} \sum\limits_{i=1}^{M_{3}}\pi_{i,M_{4}}^{(2)}\text{, }
P_{D_{5}} = \lambda_{5}\bar{\mu}_5 \pi^{(3)}_{M_{5}}\text{.} \nonumber
\end{align}
For the case of $Q_{6}$, drops can occur when $Q_{6}$ is on $(M_{6})^{\text{th}}$ or $(M_{6}-1)^{\text{th}}$ state. The drop rate of $Q_{6}$ is shown below
\begin{align}\label{eq: DrQ6Small}
P_{D_{6}}=p_{02} (1-\mu_{6}) \pi^{(4)}_{M_{6}-1}+ (p_{01} +2p_{02})\pi^{(4)}_{M_{6}}(1- \mu_{6})\text{.}
\end{align} 
The average length of each queue is given by
\begin{align}\nonumber
\bar{Q}_{1}  & = \sum\limits_{i=0}^{M_{1}}\sum\limits_{j=0}^{M_{2}} \pi_{i,j}^{(1)} i\text{, }\bar{Q}_{2} = \sum\limits_{j=0}^{M_{2}}\sum\limits_{i=0}^{M_{1}} \pi_{i,j}^{(1)} j\text{, }\bar{Q}_{3}  = \sum\limits_{i=0}^{M_{3}}\sum\limits_{j=0}^{M_{4}} \pi_{i,j}^{(2)} i\text{, }\\\nonumber
\bar{Q}_{4} &= \sum\limits_{j=0}^{M_{4}}\sum\limits_{i=0}^{M_{3}}\pi^{(2)}_{i,j} j\text{, }
\bar{Q}_{5}  = \sum\limits_{i=0}^{M_5} \pi_{i}^{(3)}i \text{, }
\bar{Q}_{6} = \sum\limits_{i=0}^{M_{6}} \pi_{i}^{(4)}i\text{.}
\end{align}
Therefore, the system drop rate and the average number of tasks in the system can be described as
$P_{D} = \sum\limits_{i\in\mathcal{K}}P_{D_{i}} \text{ and } \bar{Q} = \sum\limits_{i\in \mathcal{K}} \bar{Q}_{i}\text{,}$respectively.

\subsection{Delay and throughput analysis}
We denote by $T_{i}$ the throughput of each queue $i$. Since we consider finite buffers, and consequently packet drops, the throughput of each queue as well as system throughput depend on both the arrival  and drop rate. Therefore, the throughput for $Q_{1}-Q_{6}$ is calculated as
\begin{align}\label{eq: throughput}
T_{i} = \lambda_{i} - P_{D_{i}}\text{, } \forall i \in \mathcal{K}\text{,}
\end{align}
that is equivalent to the effective arrival rate of each queue. 

In order to derive the per packet average delay expression, we first derive the expression of packet average delay  for each queue that includes both the queueing and transmission delay. We denote by $D_{i}$, the average per packet delay for each queue. We utilize the Little's theorem \cite{BertsekasDataNetworks} and we derive the corresponding expressions as shown below
\begin{align}\label{eq: delay}
D_{i} = \frac{\bar{Q}_{i}}{T_{i}} + \frac{1}{\mu_{i}}\text{, } \forall i \in \mathcal{K}\text{.}
\end{align}
Finally, the per packet average delay is calculated as 
\begin{align}
D_{\text{sys}} = \alpha (D_{1}+D_{2}) + (1-\alpha) (D_{3}+D_{4} +D_{5}) +D_{6} \text{.}
\end{align}

\section{Scaled-Up System}
\label{Sec: ScaledUpSystem}
\begin{figure}[t!]
	\centering
	\includegraphics[scale=0.35]{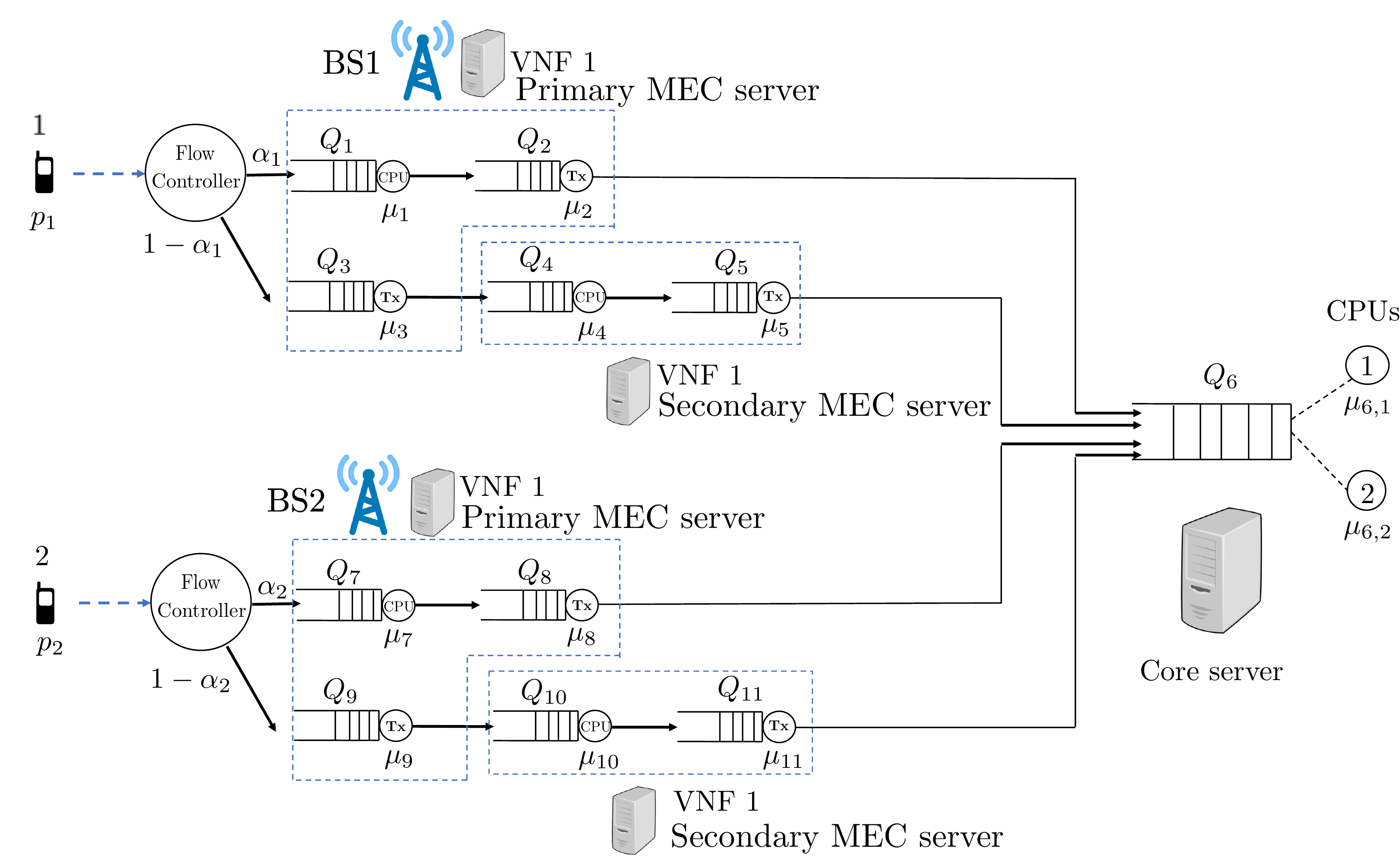}
	\caption{Two user devices transmit tasks to two primary MEC servers (locating at base stations) respectively, assuming that there are two CPUs in the core server.}
	\label{Fig: ScaledUpSystem}
\end{figure}

In this section, we explain in details how the analysis that is shown above can be used in general cases of multiple user devices communicating to multiple base stations that are connected to the core server. We provide the analysis and methodology for two base stations. The same methodology can be applied for a system with arbitrary number of base stations. However, we choose the case of two base stations in order to facilitate the presentation. We analyze the system that is shown in Fig. \ref{Fig: ScaledUpSystem}, by following the same methodology in order to derive the approximate analytical expressions: we decompose the system into subsystems and analyze each subsystem independently. Furthermore, we consider that the server in the core has two CPUs. Therefore, the number of departures is up to two tasks per time slot. The number of arrivals can be up to four tasks per time slots. We decompose the system into seven individual subsystems: 1) $Q_{1}$ and $Q_{2}$, 2) $Q_{3}$ and $Q_{4}$, 3) $Q_{5}$, 4) $Q_{6}$, 5) $Q_{7}$ and $Q_{8}$, 6) $Q_{9}$ and $Q_{10}$, and 7) $Q_{11}$. For the subsystems $1-3$ and $5-7$, we apply the analysis that is shown in Sections \ref{Sec: PerAnalysis}-\ref{Sec: KPM}. For Subsystem 4, we provide the structure of the transition matrix in below. First, we define the arrival rate of  $Q_{6}$ that depends on the throughput of $Q_{2}$, $Q_{5}$, $Q_{8}$, and $Q_{11}$. The arrival rates of $Q_{6}$ that occur due to $Q_{2}$, $Q_{5}$, $Q_{8}$, and $Q_{11}$ are $\lambda_{6,2}  = \text{Pr}\left\{Q_{2}>0\right\}\mu_{2}\text{, }$ $\lambda_{6,5}    = \text{Pr}\left\{Q_{5}>0\right\} \mu_{5} \text{,}$ $\lambda_{6,8} = \text{Pr}\left\{Q_{8}>0\right\} \mu_{8} \text{, and } \lambda_{6,11}  = \text{Pr}\left\{Q_{11}>0\right\} \mu_{11} \text{,}$
respectively.
Furthermore, we assume that the service time  are identically geometrically distributed. Therefore, $\mu_{6,1}=\mu_{6,2}$ and we set $\mu_{6,1}=\mu_{6,2}=\mu$. Let $X$ be a random variable that represents the number of departures at each time slot.
Given that the queue is non-empty, $X$ follows a binomial distribution with success probability $\mu$. The Probability Mass Function (PMF) is given by $\text{Pr} \left\{X=k\right\} = {n \choose k} \mu^{k} (1-\mu)^{n-k} \text{,}$
where $n$ is the number of cores, and $k$ is the number of departures in one slot. In our case, $n=2$, and $k\leq 2$.
Therefore, we can easily calculate the probabilities to have $k$ departures at each time slot.
We model this subsystem as one Markov chain. The corresponding transition matrix is shown below
\begin{small}
	\begin{align}\label{Eq: TranCore}
	\mathbf{P}_{6}=
	\left[\begin{array}{ccccccccccc}
	p_{00} & p_{01} & p_{02} & p_{03}   &   p_{04}   &  &  &  &  &  &  \\
	p_{10}  & p_{11} & p_{12}  &  p_{13}  &    p_{14}  & p_{15} &   &   &   &    &    \\
	b_{6}   & b_{5}  & b_{4}    & b_{3}     &   b_{2}    & b_{1} & b_{0} &   &   &   &  \\
	&    \ddots       & \ddots & \ddots    &  \ddots   &  \ddots & \ddots  & \ddots &  &  & \\
	&  &   b_{6}  & b_{5}   & b_{4}      &   b_{3}    & b_{2} & b_{1} & b_{0}  &  & \\ 
	&  &             &  b_{6}  & b_{5}      &   b_{4}  	& b_{3} & b_{2} & b_{1} & b_{0}\\
	&  &             &            &  b_{6}    &    b_{5}    & b_{4} & b_{3} & b_{2} & b_{0} + b_{1} \\
	&  &             &            &            &  b_{6}    &    b_{5}    & b_{4} & b_{3}  & b_{2} + b_{1} + b_{0} \\
	&  &             &  & &            &  b_{6}    &    b_{5}    & b_{4}  & b_{3}+b_{2}+b_{1}+b_{0} \\
	&  &             &  & & &          &  b_{6}    &    b_{5}  &  b_{4} + b_{3}+b_{2}+b_{1}+b_{0} \\
	\end{array}\right]\text{,}
	\end{align}
\end{small}
where the elements of the matrix are calculated in  Appendix A.

Let $i$ and $j$ be the $i$-th row and $j$-th column of the transition matrix in (\ref{Eq: TranCore}), respectively.
We observe that the transition matrix has repeated elements from the  third row to the $(M_{6}-A_{\text{max}} + 1)^{\text{th}}$ row, where $A_{\text{max}}$ is the maximum number of tasks that can arrive in $Q_{6}$ at each time slot. Therefore for the system above, $A_{\text{max}}=4$. Thus, after calculating the transition probabilities for the first three rows, it is easy to construct the transition matrix without calculating each element independently.

In order to derive analytical expressions for  key performance metrics, we need to calculate the steady state distribution of the Markov chain of $Q_{6}$. We apply EVD  to derive the steady state distribution and follow the same methodology as in Section \ref{Sec: PerAnalysis}.

After obtaining the steady state distribution, we can calculate the drop rate of $Q_{6}$. In this case, at each time slot, more than one task can be dropped. The reason is because more than one task can arrive at each time slot. Drops occur when $Q_{6}$ has less than four vacant positions. This corresponds to the $(M_{6}-3)^{\text{th}}$ state. To summarize, drops occur when $Q_{6}$ is on the $(M_{6}-3)^{\text{th}}$, $(M_{6}-2)^{\text{th}}$, $(M_{6}-1)^{\text{th}}$, or $(M_{6})^{\text{th}}$ state. Below, we calculate the drop rate given that $Q_{6}$ is on a particular state
\begin{align}\nonumber
P_{D_6|Q_{6}=M_{6}}    & =p_{01}\pi_{M_6}^{(4)}(1-\mu)^2+2p_{02}\pi_{M6}^{(4)}(1-\mu)^2+p_{02}\pi_{M_6}^{(4)}2\mu(1-\mu)+3p_{03}\pi_{M6}^{(4)}(1-\mu)^2\\\nonumber
&+2p_{03}\pi_{M_6}^{(4)}2\mu(1-\mu)+p_{03}\pi_{M_{6}}^{(4)}\mu^{2}+4p_{04}\pi_{M_{6}}^{(4)}(1-\mu)^2+3p_{04}\pi_{M_{6}}^{(4)}2(1-\mu)\mu\\
& +2p_{04}\pi_{M_{6}}^{(4)}\mu^2\text{,} \\\nonumber
P_{D_6|Q_{6}=M_{6}-1}  & =	p_{02}\pi_{M_6-1}^{(4)}(1-\mu)^2+2p_{03}\pi_{M_{6}-1}^{(4)}(1-\mu)^2+p_{03}\pi_{M_{6}-1}^{(4)}2(1-\mu)\mu\\
&+3p_{04}\pi_{M_{6}-1}^{(4)}(1-\mu)^2+2p_{04}\pi_{M_{6}-1}^{(4)}2(1-\mu)\mu+p_{04}\pi_{M_{6}-1}^{(4)}\mu^2\text{,}\\
P_{D_6|Q_{6}=M_{6}-2} & =  	p_{03}\pi_{M_{6}-2}^{(4)}(1-\mu)^2+2p_{04}\pi_{M_{6}-2}^{(4)}2(1-\mu)\mu+p_{04}\pi_{M_{6}-2}^{(4)}\mu^2\text{,}\\
P_{D_6|Q_{6}=M_{6}-3} & = p_{04}\pi_{M_{6}-3}^{(4)}(1-\mu)^2\text{.}
\end{align}
Finally, the total drop rate of $Q_{6}$ is shown below
\begin{align}
P_{D_6} = P_{D_6|Q_{6}=M_{6}-3} + P_{D_6|Q_{6}=M_{6}-2}  + P_{D_6|Q_{6}=M_{6}-1}  + P_{D_6|Q_{6}=M_{6}}  \text{.}
\end{align}

We derive the throughput and delay for each queue by using the equations in (\ref{eq: throughput}), (\ref{eq: delay}).
The average delay that arrives in the first and the second base station are described below 
\begin{align}
D_{BS_1}  & =\underbrace{\alpha_{1}(D_{1} + D_{2}) +(1-\alpha_{1})(D_{3} + D_{4} + D_{5})}_{A_{1}} + D_{6}\text{,}\\
D_{BS_2} & = \underbrace{\alpha_{2} (D_{7} + D_{8}) + (1-\alpha_{2})(D_{9} + D_{10} + D_{11})}_{A_{2}} + D_{6}\text{,} 
\end{align}
respectively. Finally, the per packet average delay is calculated as
\begin{align}
D_{sys} = p_{1} A_{1} + p_{2} A_{2} + D_{6}\text{.}
\end{align}

\section{Numerical and Simulation Results}\label{Sec:Sim} 
In this section, we evaluate the performance of our approximated model by comparing the analytical and simulation results. 
First, we study the performance of the system with one base station by showing how different parameters and routing decisions affect the system performance in terms of delay, throughput, and drop rate. Second, we  study the performance of a system with two base stations. Our goal is to evaluate the performance of our approximation model for larger systems and provide insights on the operation of the considered setup. We developed a MATLAB-based behavioural simulator and each case run for $10^6$ timeslots.

\subsection{The one base station case}
\subsubsection{Effect of the routing decision on the system performance - analytical vs simulation results}
\begin{figure}[t!]
	\centering
	\subfloat[$\mu_{1}=\mu_{2}=0.2$, $\mu_{3}=\mu_{4}=\mu_{5}=0.6$.]{%
		\includegraphics[width=0.49\linewidth]{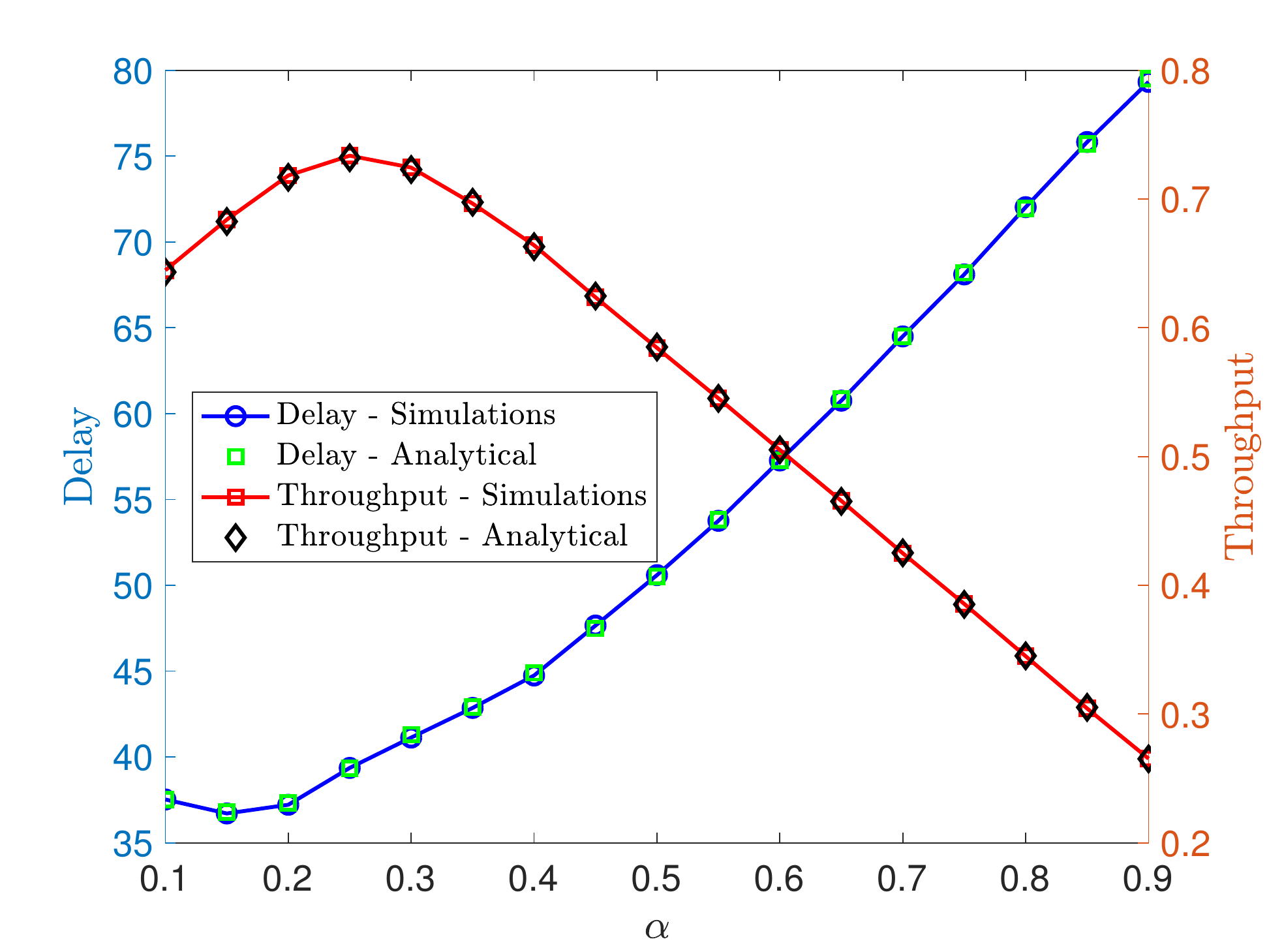}
		\label{Fig. RoutingSmallmu1}}
	\hfill
	\subfloat[$\mu_{1}=\mu_{2}=0.6$, $\mu_{3}=\mu_{4}=\mu_{5}=0.2$.]{%
		\includegraphics[width=0.49\linewidth]{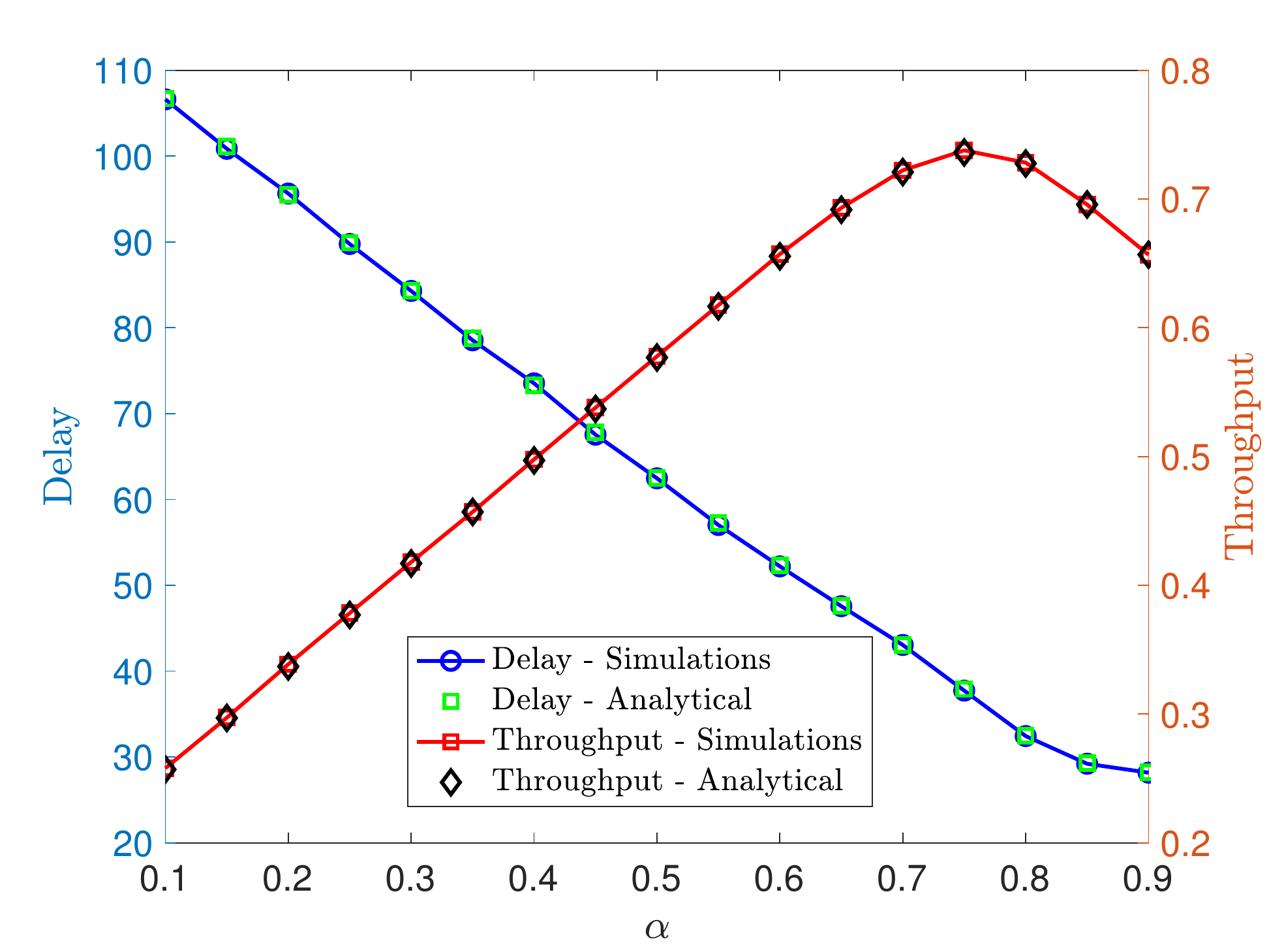}
		\label{Fig. RoutingSmallmu3}
	}
	\caption{Effect of routing decision on the system performance. $\mu_{6}=1$, $p=0.8$, $M_{i}=10$ for $1\leq i \leq 5$, $M_{6}=100$.}
	\label{Fig. RoutingDiffSetUp}
\end{figure}
\begin{figure}[t!]
	\centering
	\includegraphics[scale=0.45]{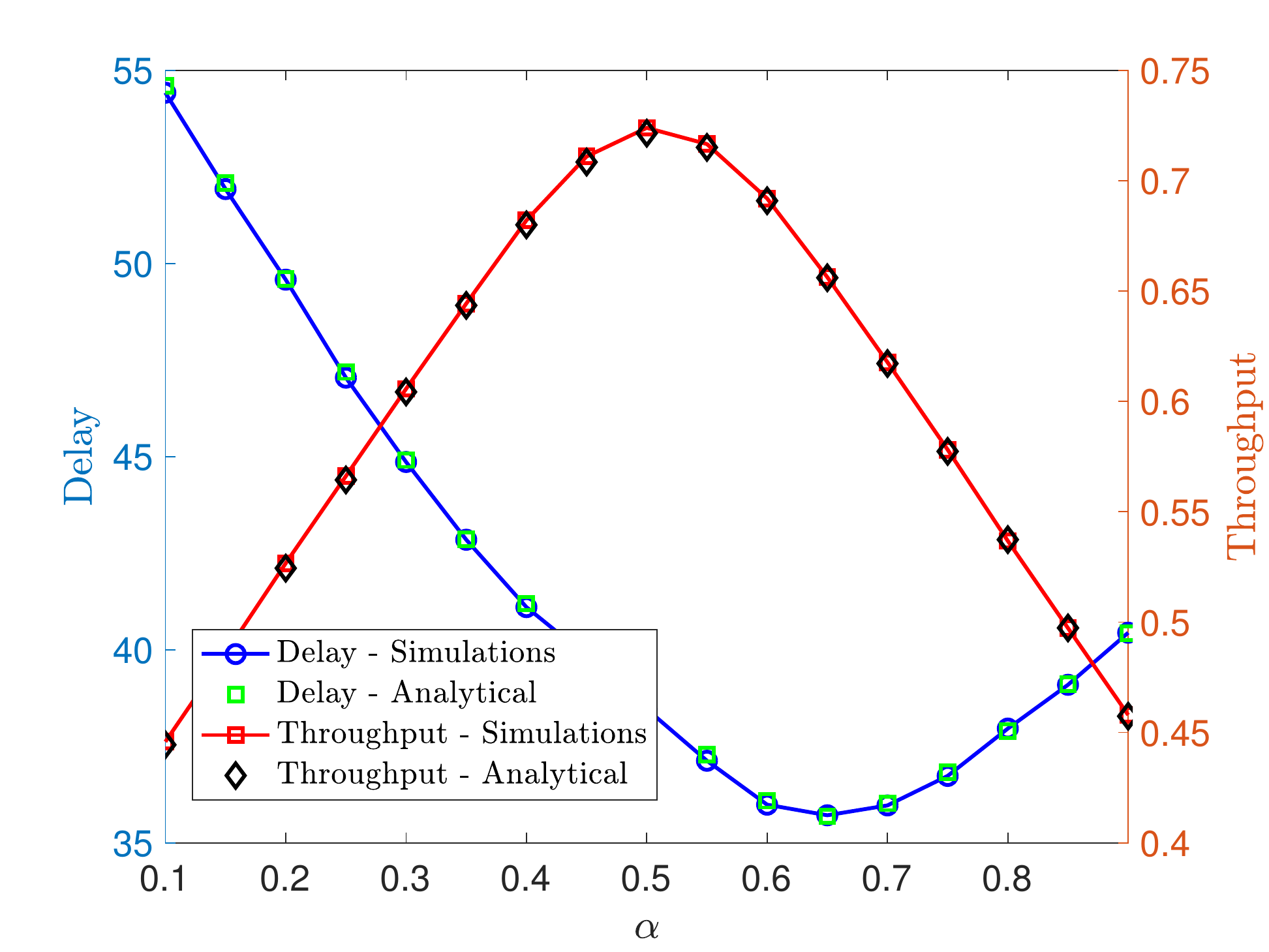}
	\caption{Effect of routing decision on the system performance. $\mu_i = 0.5$ for $1 \leq i \leq 5$ , $\mu_{6}=1$, $p=0.8$, $M_{i}=10$ for $1\leq i \leq 5$, $M_{6}=100$.}
	\label{Fig: RoutingEq} 
\end{figure}
In Fig. \ref{Fig. RoutingDiffSetUp} and Fig. \ref{Fig: RoutingEq}, we provide results that show how different routing decisions affect the performance of the system  for different settings of parameters $\mu_i, i = 1, 2, \ldots, 5$ . We also compare the simulation results with those of analytical in order to evaluate the performance of our approximated model. In the first case depicted in Fig. \ref{Fig. RoutingSmallmu1}, the capabilities of the MEC Server 1 in terms of service rates are lower than those of the MEC server 2. We observe that the optimal values of throughput and delay are achieved for values of $\alpha$ that are close to $0.2$.  The value of $\alpha$ that minimizes delay, it does not necessarily maximize the throughput. For larger values of drop rates, we may have shorter delay of the system but smaller throughput as well. In Fig. \ref{Fig. RoutingDiffSetUp}, we observe results for smaller capabilities of the MEC server $2$. 
In Fig. \ref{Fig: RoutingEq}, the capabilities of the MEC servers are identical. We observe that the maximum throughput is achieved for $\alpha=0.5$. However, the optimal $\alpha$ in terms of end-to-end  delay is different. The value of $\alpha$ that minimizes the delay is around $0.65$. The flow controller forwards larger amount of traffic to the first flow. This strategy minimizes the delay because the tasks have to traverse a smaller amount of queues and therefore, face shorter waiting times. However, this has an impact on the system performance in terms of drop rate and therefore, the system throughput.

\subsubsection{Effect of $\mu_{1}$ and $\mu_{3}$  on the system throughput systems with small buffer size}
\begin{figure}[t!]
	\centering
	\subfloat[Optimal values of $\alpha$.]{%
		\includegraphics[width=0.32\linewidth]{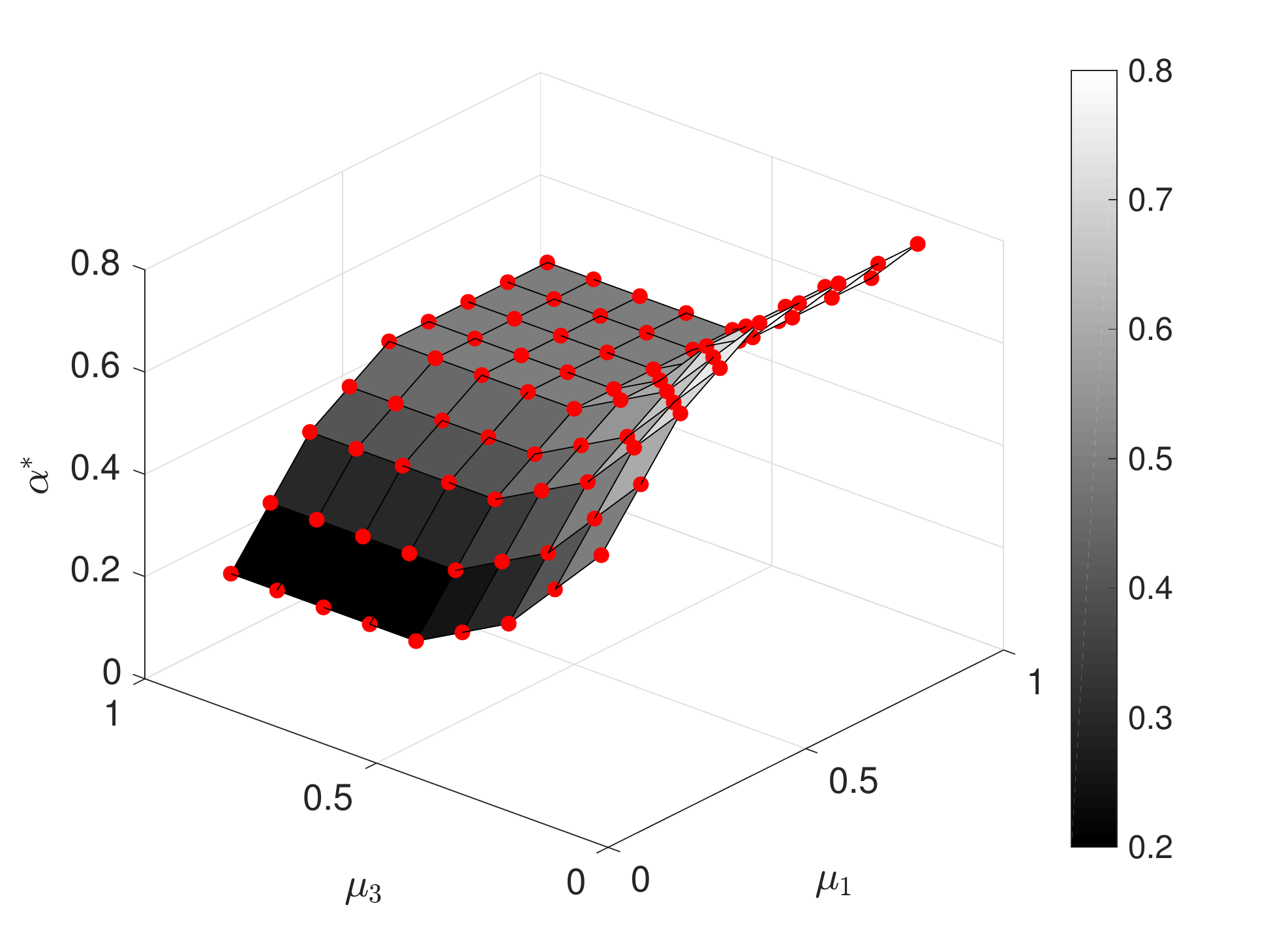}
		\label{Fig: OptalphaThr}}
	\hfill
	\subfloat[Achieved maximum throughput.]{%
		\includegraphics[width=0.32\linewidth]{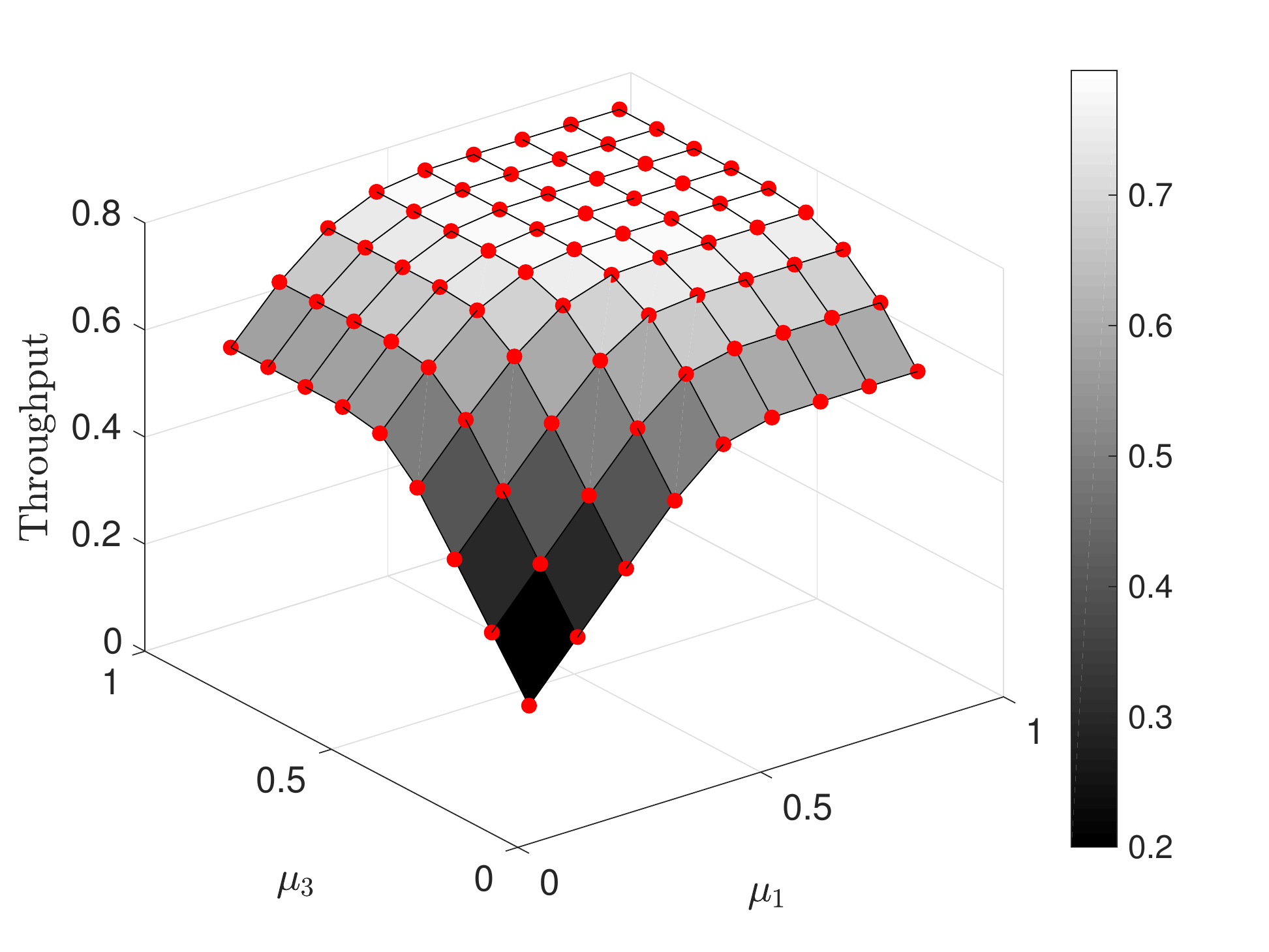}
		\label{Fig: MaxThr}
	}
	\hfill
	\subfloat[System delay.]{%
		\includegraphics[width=0.32\linewidth]{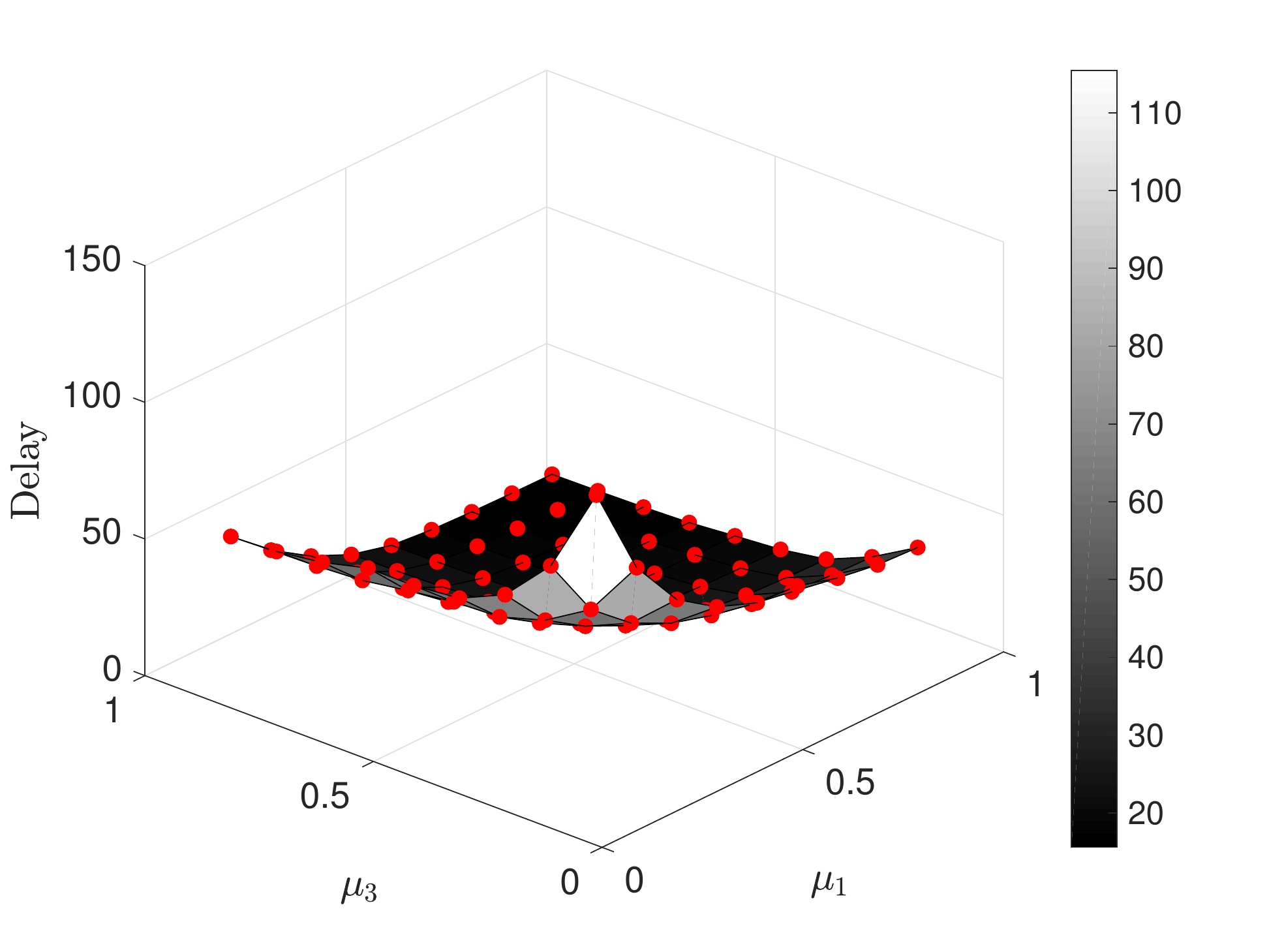}
		\label{Fig: Delaythroptmu1mu3}
	}
	\caption{Objective: To maximize the system throughput. $\mu_{2}=\mu_{4}=\mu_{5}=0.5$, $\mu_{6}=0.95$, $p=0.8$, $M_{i}=10$ for $1\leq i \leq 5$, $M_{6}=100$.}
	\label{Fig: Throptmu1mu3}
\end{figure}
In this subsection, we provide results for the performance of the system in terms of throughput. In this case, our objective is to maximize the throughput in small buffered systems. In Fig. \ref{Fig: Throptmu1mu3}, we provide results that show the optimal routing decisions and the corresponding achieved throughput. The service rates of $Q_{2}$, $Q_{4}$ and $Q_{5}$ are fixed. We find the optimal routing decisions by applying exhaustive search for different values of $\mu_{1}$ and $\mu_{3}$. In Fig. \ref{Fig: OptalphaThr}, we show the optimal routing decisions for different values of $\mu_{3}$ and $\mu_{1}$. We observe that when $\mu_{1}<\mu_{3}$, the optimal value of $\alpha$ is close to $0.18$. Therefore, almost $80 \%$ of the traffic is processed by the MEC server $2$. On the other hand, when $\mu_{1}>\mu_{3}$, the largest part of the traffic is handled by the MEC server $1$. For the cases of $\mu_{1}=\mu_{3}$, the traffic is splitted between the two flows and the router achieves a balance between them. In Fig. \ref{Fig: MaxThr}, we provide results that show the optimal achieved throughput for different values of $\mu_{1}$ and $\mu_{3}$. Note that the maximum achievable throughput is equal to the arrival rate, i.e., $0.8$. However, the maximum  throughput is not achieved for this setup because of limited buffer capacity. We observe that even for large values of the service rates, the throughput is close to the arrival rate. This indicates  that buffers with higher capacities are required in order to achieve a higher maximum throughput.
\subsubsection{Effect of $\mu_{1}$ and $\mu_{3}$ on the system delay in systems with large buffer size}
\begin{figure}[t!]
	\centering
	\subfloat[Optimal values of $\alpha$.]{%
		\includegraphics[width=0.31\linewidth]{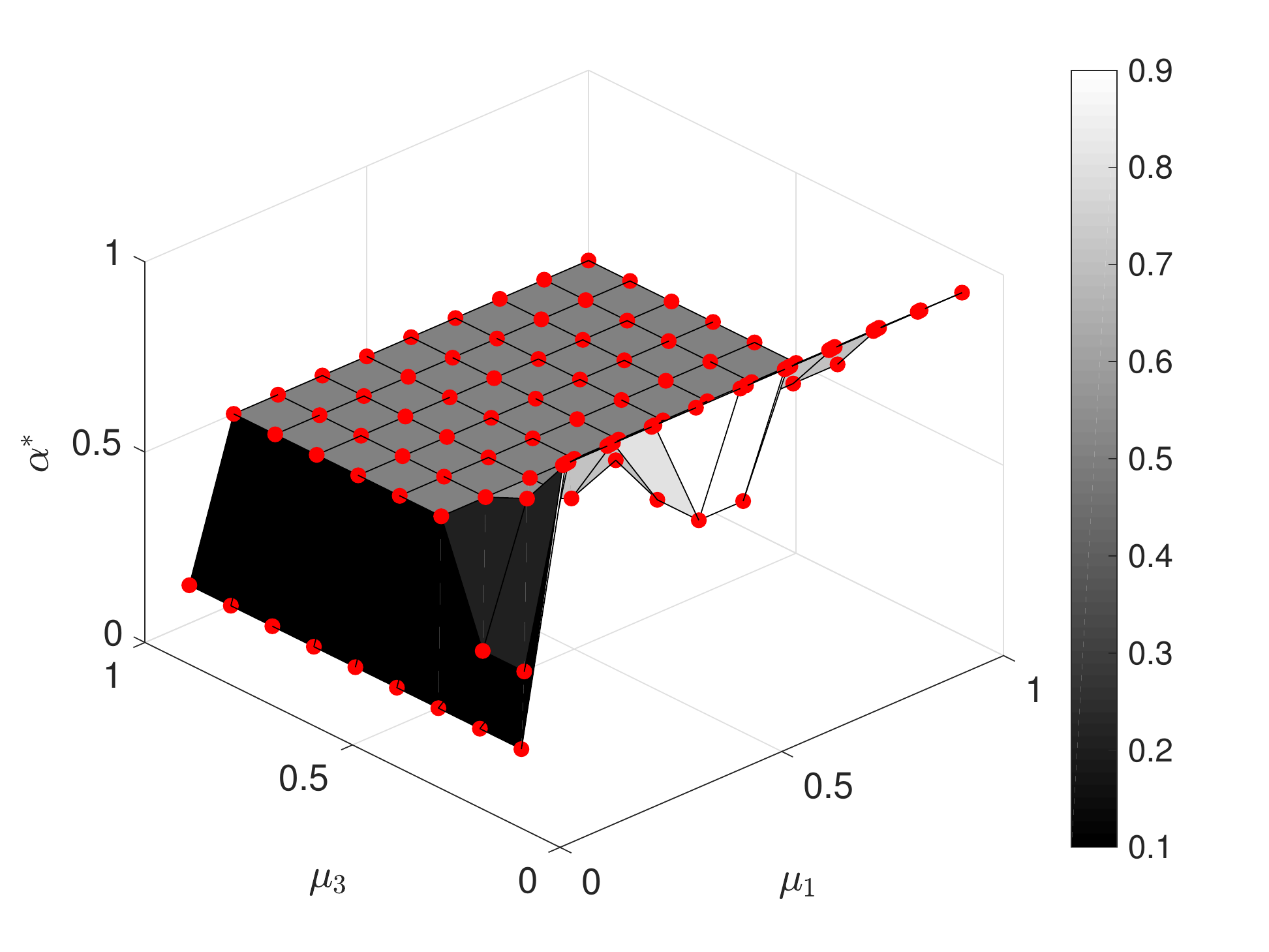}
		\label{Fig: OptalphaDel}}
	\hfill
	\subfloat[Achieved minimum delay.]{%
		\includegraphics[width=0.31\linewidth]{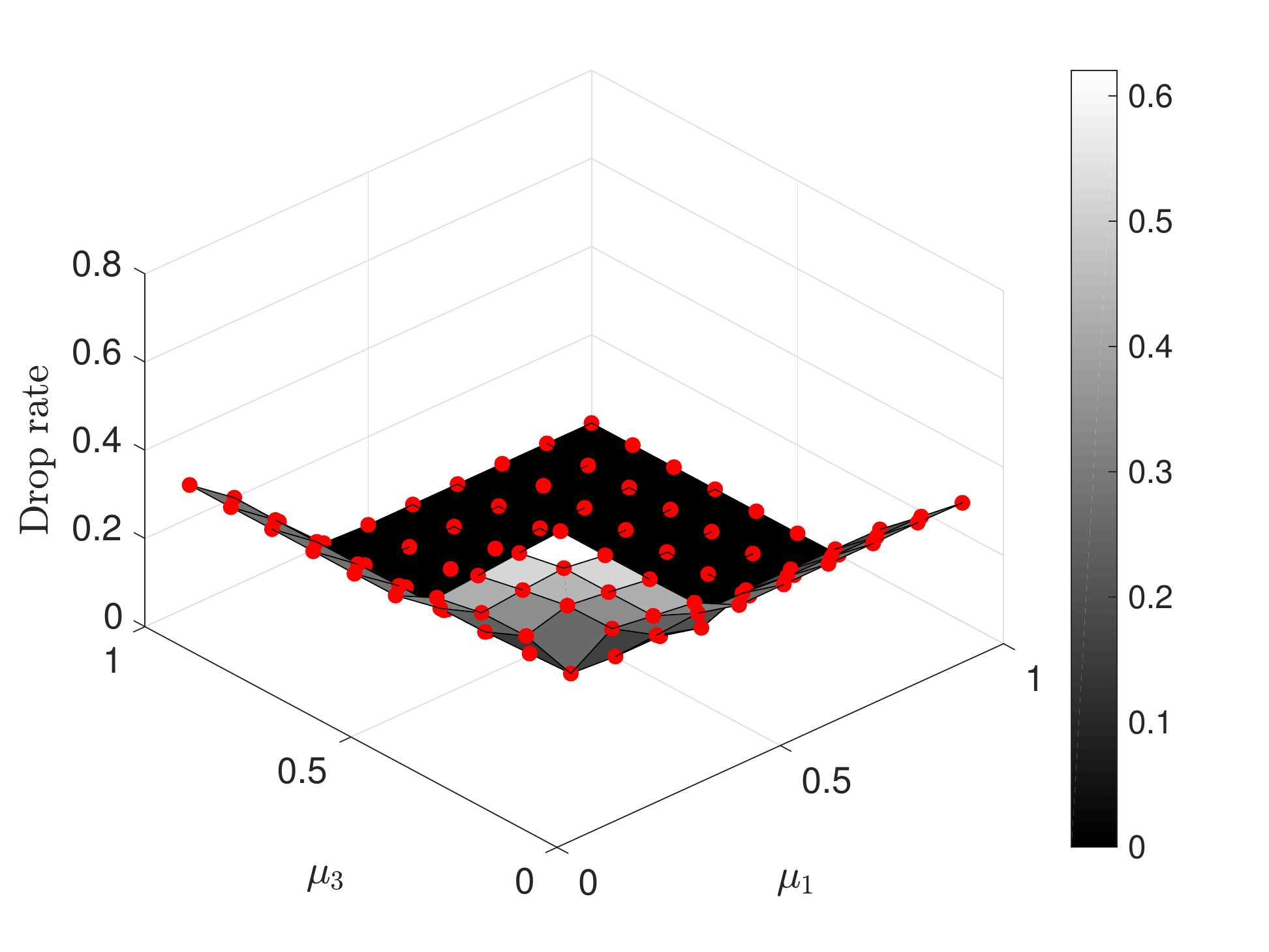}
		\label{Fig: MinDelay}
	}
	\hfill
	\subfloat[System throughput.]{%
		\includegraphics[width=0.31\linewidth]{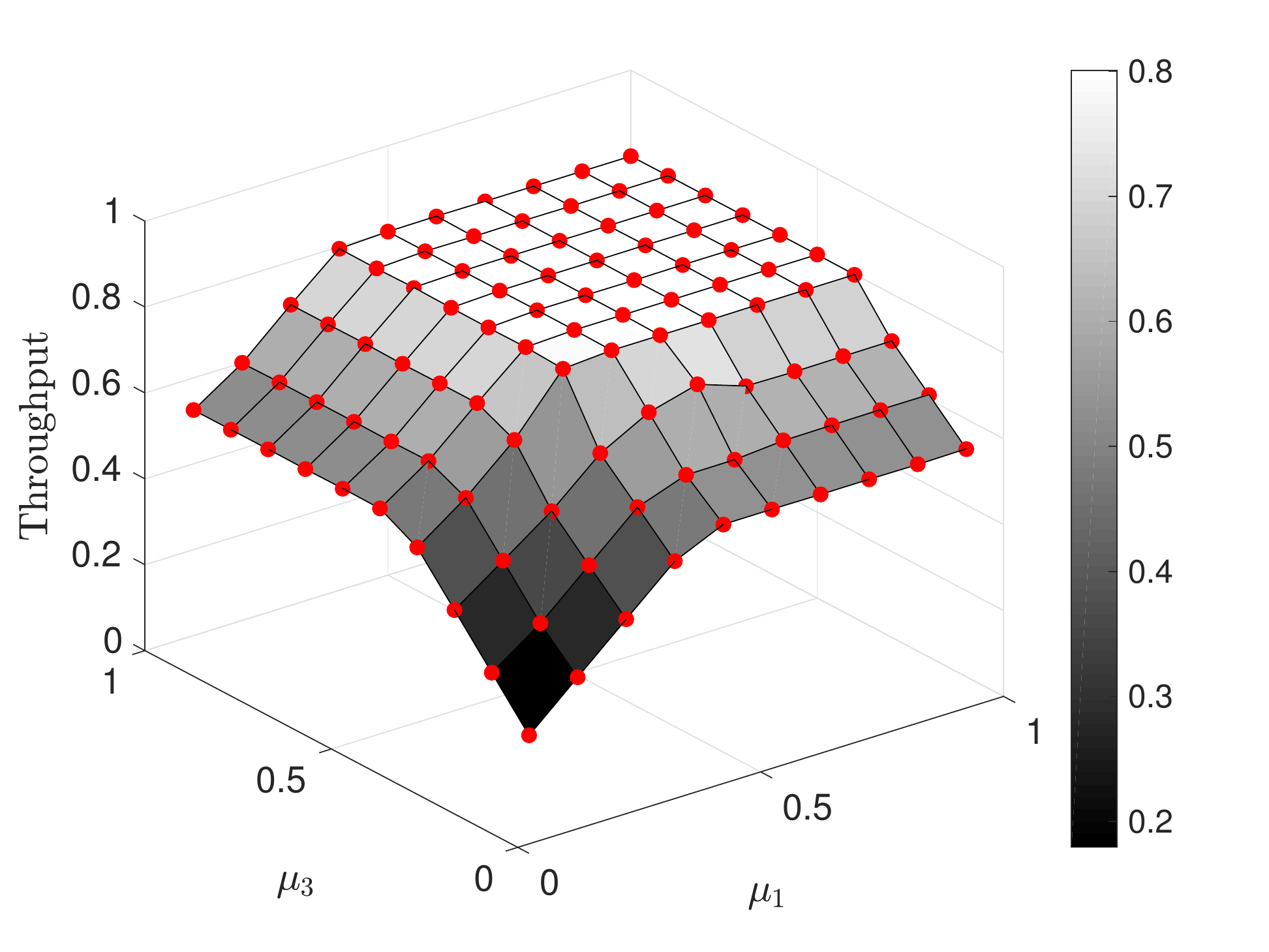}
		\label{Fig: MinDelayThrou}
	}
	\caption{Objective: To minimize the system delay. $\mu_{2}=\mu_{4}=\mu_{5}=0.5$, $\mu_{6}=0.95$, $p=0.8$, $M_{i}=50$ for $1\leq i \leq 5$, $M_{6}=100$.}
	\label{Fig: Delayoptmu1mu3}
\end{figure}

In this subsection, we provide results that show the performance of the system in terms of delay. In this case, we minimize the delay in systems with large buffer size by selecting the optimal values of $\alpha$ as shown in Fig. \ref{Fig: Delayoptmu1mu3}. In addition, the optimal values of $\alpha$ are depicted in Fig. \ref{Fig: OptalphaDel}. We observe that when $\mu_{1}=\mu_{3}$, the optimal routing decision is to split the traffic into the two flows equally, i.e., $\alpha=0.5$. Another interesting observation is that for even small values of $\mu_{1}$ and large values of $\mu_{3}$, the router decides to split the traffic  between the two flows. There are two reasons that explain this phenomenon. The first is that even if the capabilities of one MEC server are low, they should be still utilized in order to increase the performance. The second is that, for the case of  minimization of the delay, the router decides to send part of the traffic to the first flow because of the less number of queues  facing shorter delay. 
\subsubsection{Throughput - Delay - Drop rate trade-off}
\begin{figure}[t!]
	\centering
	\subfloat[Performance region. \label{Fig: Rega}]{%
		\includegraphics[width=0.49\linewidth]{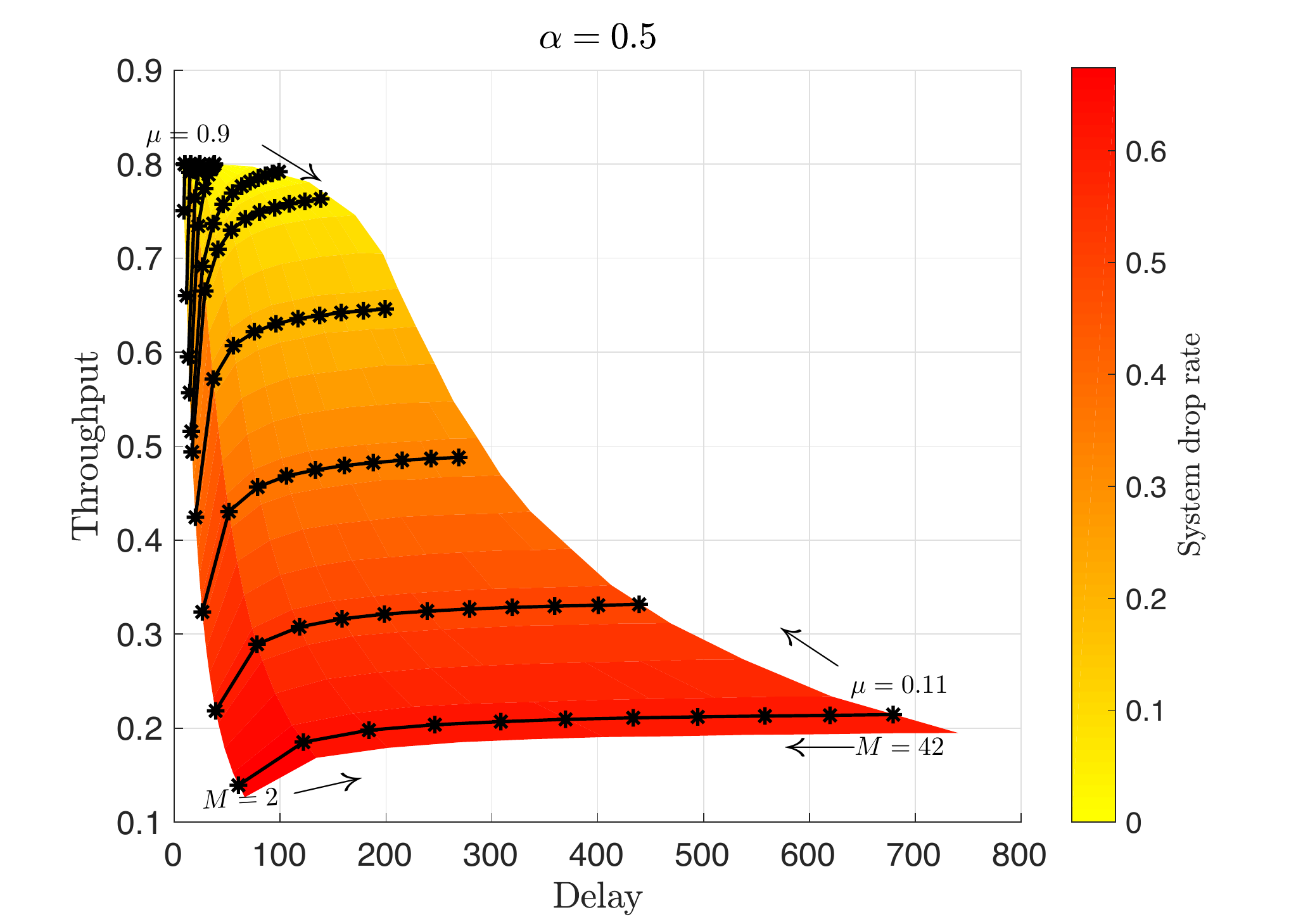}
		\label{Fig: AlphaOptimalDropExp10}}
	\hfill
	\subfloat[Performance region of our interest. \label{Fig: Regb}]{%
		\includegraphics[width=0.49\linewidth]{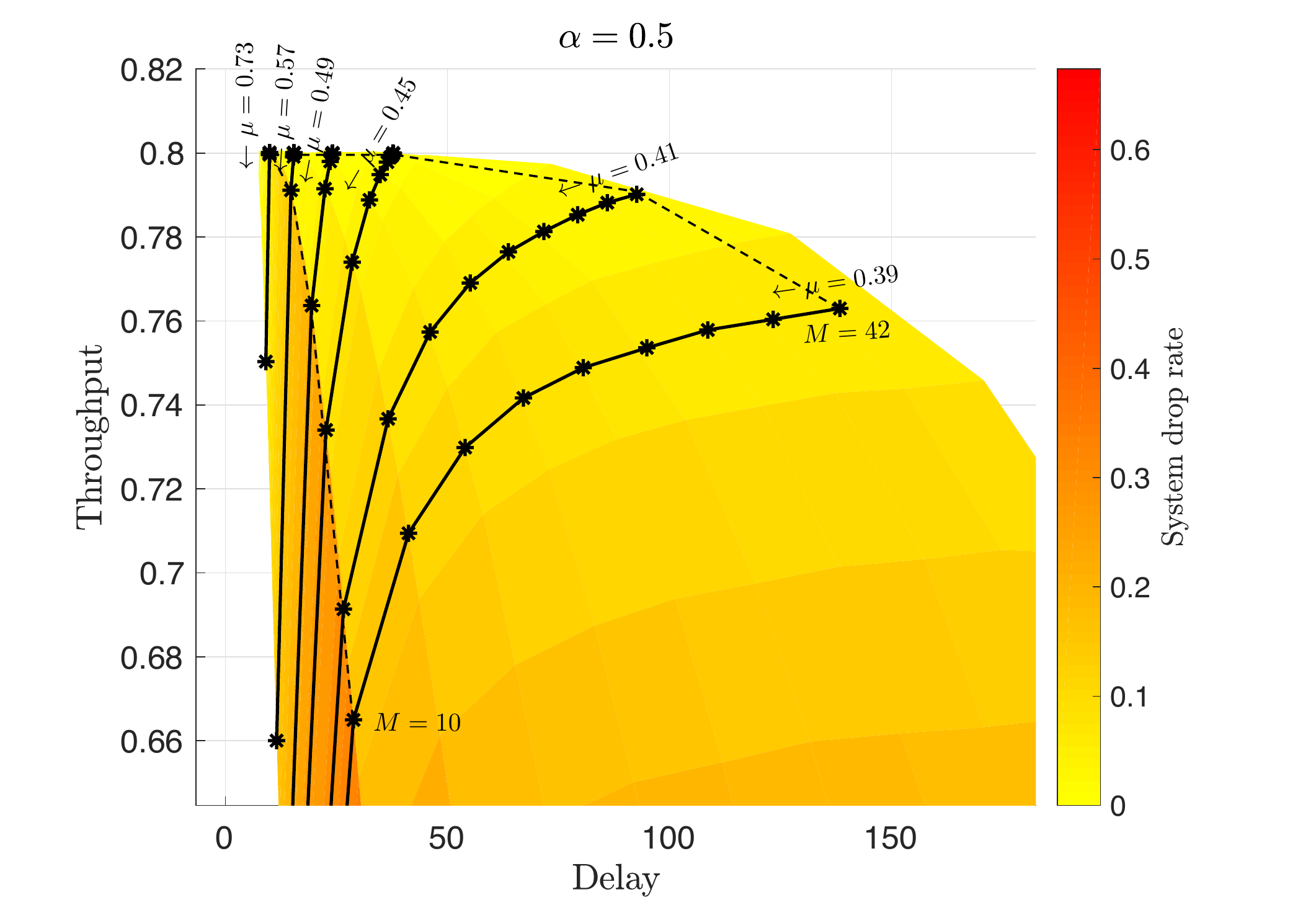}
		\label{Fig: OptimalDropExp10}
	}
	\caption{Performance region. $\mu_{3}=\mu_{4}=\mu_{5}=\mu$, $\mu_{6}=1$, $p=0.8$. $M_{i}=M$ for $1\leq i \leq 5$, $M_{6}=100$.}
	\label{Fig: Region}
\end{figure}

In this subsection, we provide results that show the performance region of the system in terms of throughput, delay, and system drop rate.
In Fig. \ref{Fig: Rega}, the performance region is shown. We generate the region by selecting different parameters of the system.
In particular, we vary the service rates and the capacities of $Q_{1}-Q_{5}$, and we take the corresponding points as shown in the diagram.
Each horizontal line is created by fixing the service rates and varying the capacity of the buffers. The color represents the system drop rate for each different case. We observe that the capacity of the buffers does not affect the system performance in terms of throughput and drop rate, when the service rates are small. On the other hand, small variations of the service rates have a significant impact on the performance metrics. In Fig. \ref{Fig: Regb}, part of the region is provided. For specific requirements of the system, for example, low latency, high reliability, or high throughput, we can provide the proper setup by utilizing the information given by the performance region figures.
\subsection{Scaled-up system}
\begin{figure}[t!]
	\centering
	\includegraphics[scale=0.5]{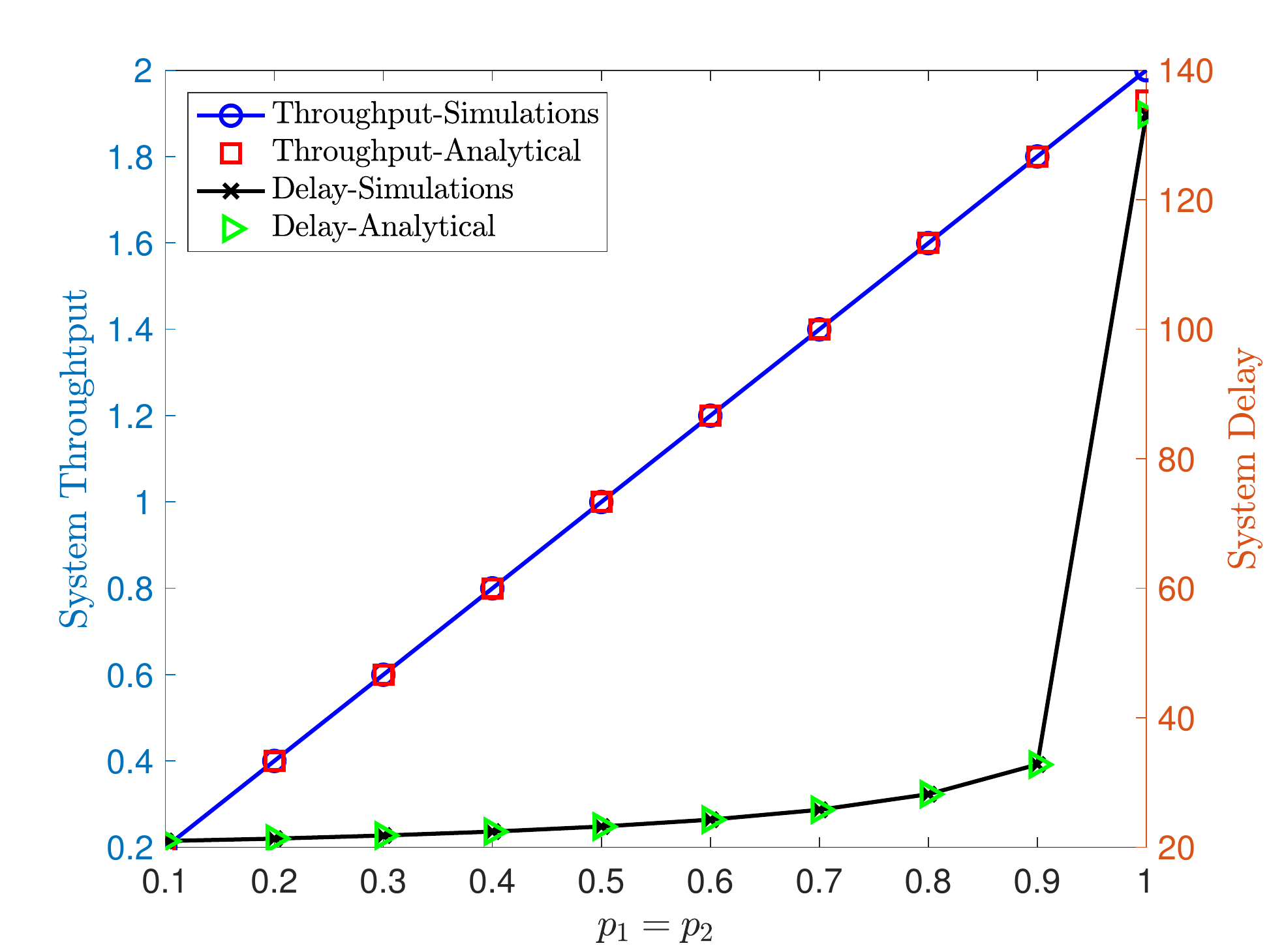}
	\caption{System throughput vs system delay. $\mu_{i}=0.6$ for $ 1\leq i \leq 5$ and $ 7\leq i \leq 11$. $\mu=1$. $M_{6}=100$, $M_{i}=50$,  for $ 1\leq i \leq 5$ and $ 7\leq i \leq 11$.}
	\label{Fig: ScaledUpEqualChanges} 
\end{figure}
\begin{figure}[t!]
	\centering
	\subfloat[Performance of BS1.]{%
		\includegraphics[width=0.49\linewidth]{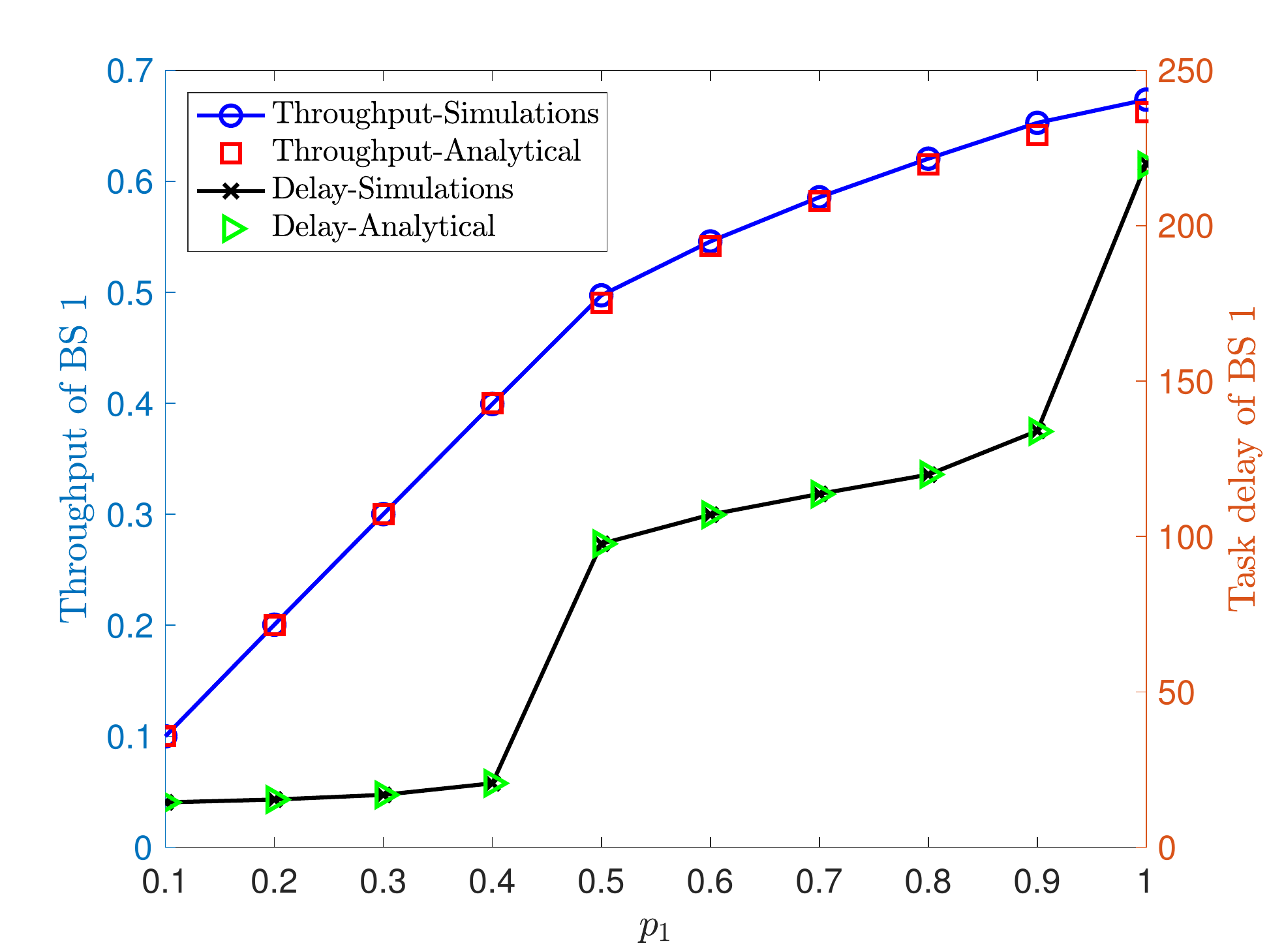}
		\label{Fig: ScaledUpVarp1BS1}}
	\hfill
	\subfloat[Performance of BS2 and the effect of high traffic of BS1 to BS2.]{%
		\includegraphics[width=0.49\linewidth]{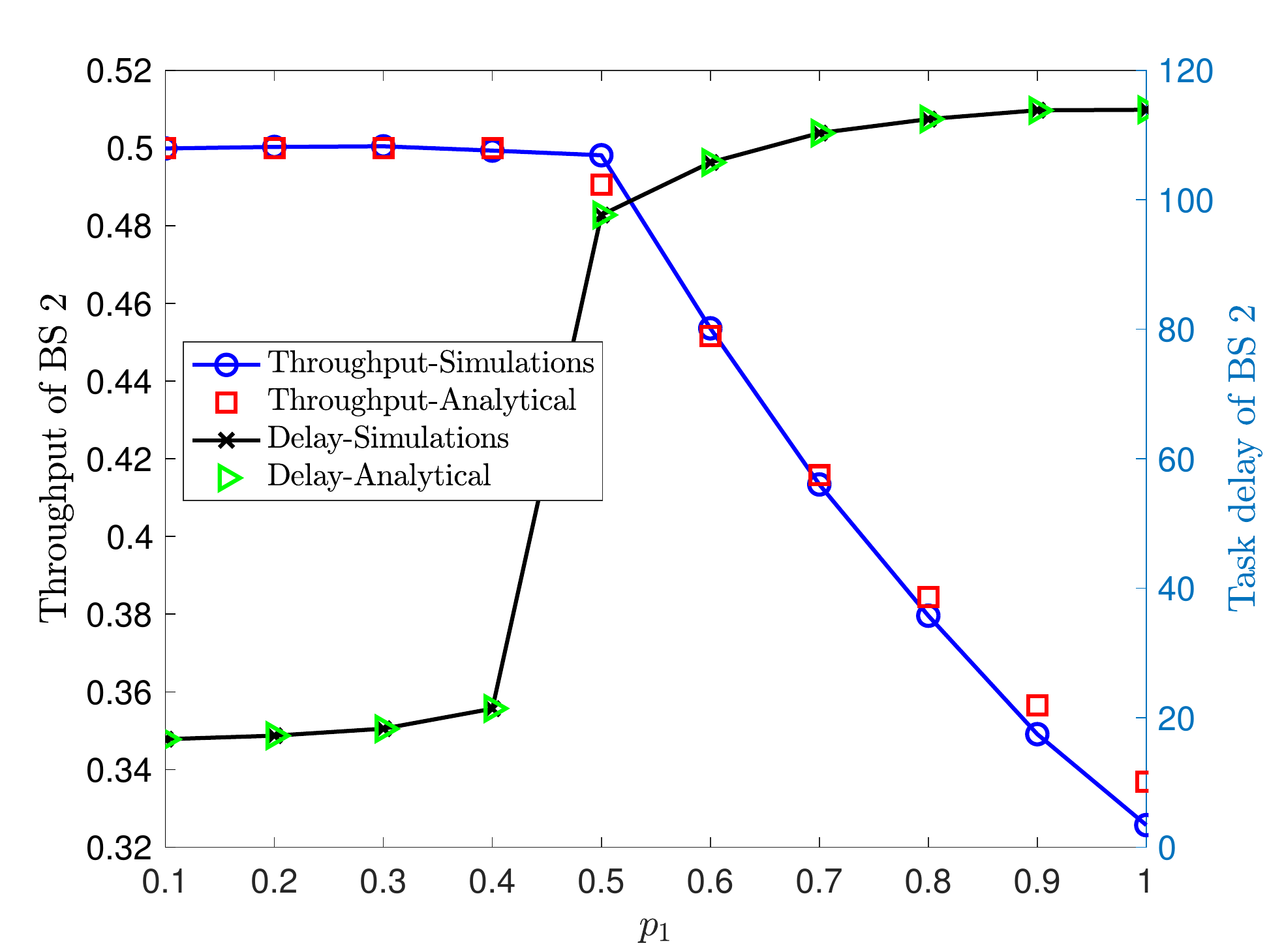}
		\label{Fig:  ScaledUpVarp1BS2}
	}
	\caption{Throughput vs delay performance. $p_{2}=0.5$, $\mu_{i}=0.5$, $\forall i$, $M_{i}=50$ for $1\leq i \leq 5$ and $ 7\leq i \leq 11$. $M_{6}=100$.}
	\label{Fig: ScaledUpVarp1}
\end{figure}

\begin{figure}[t!]
	\centering
	\subfloat[Performance of BS1.]{%
		\includegraphics[width=0.49\linewidth]{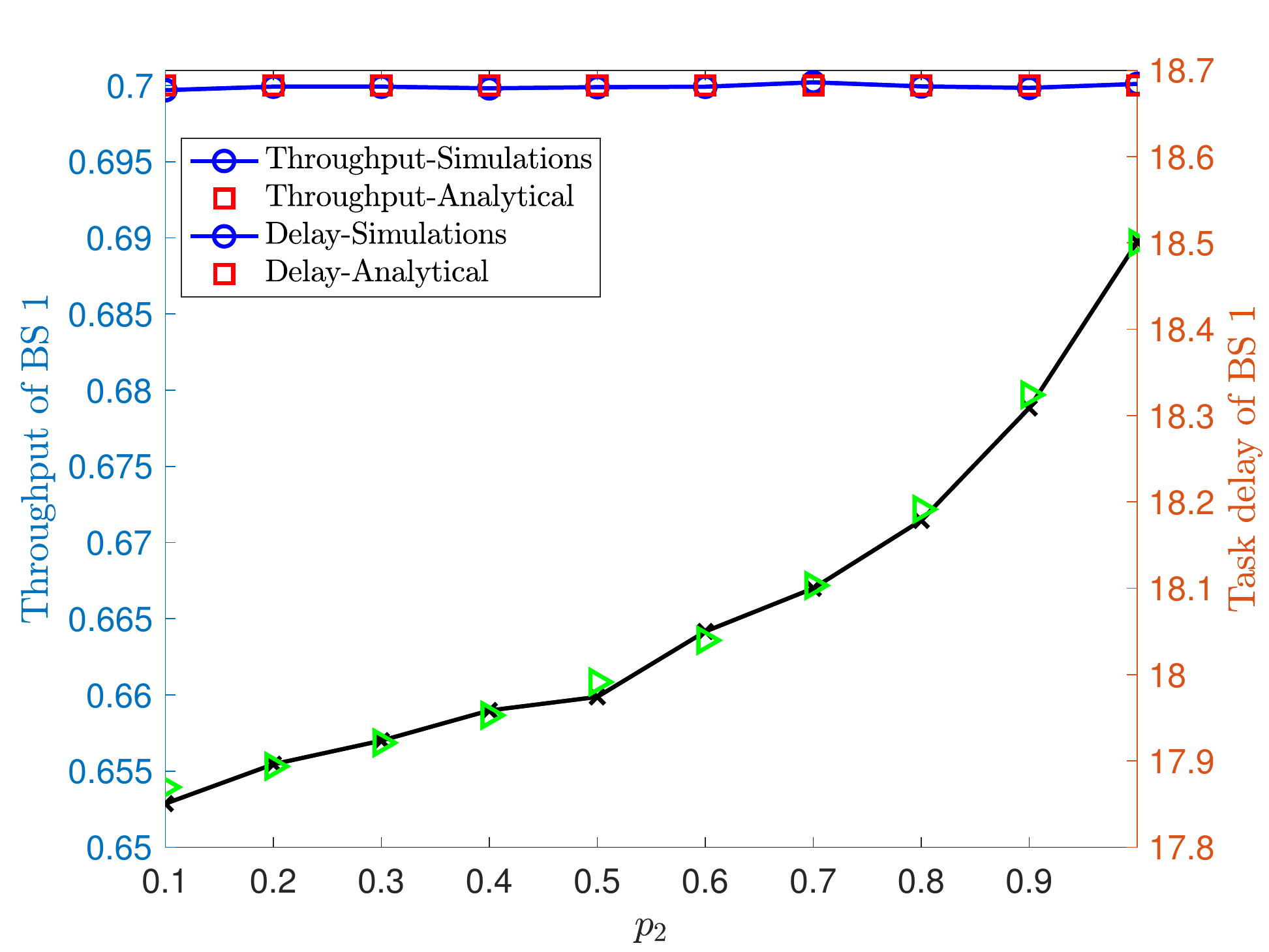}
		\label{Fig: ScaledUpVarp2BS1}}
	\hfill
	\subfloat[Performance of BS2.]{%
		\includegraphics[width=0.49\linewidth]{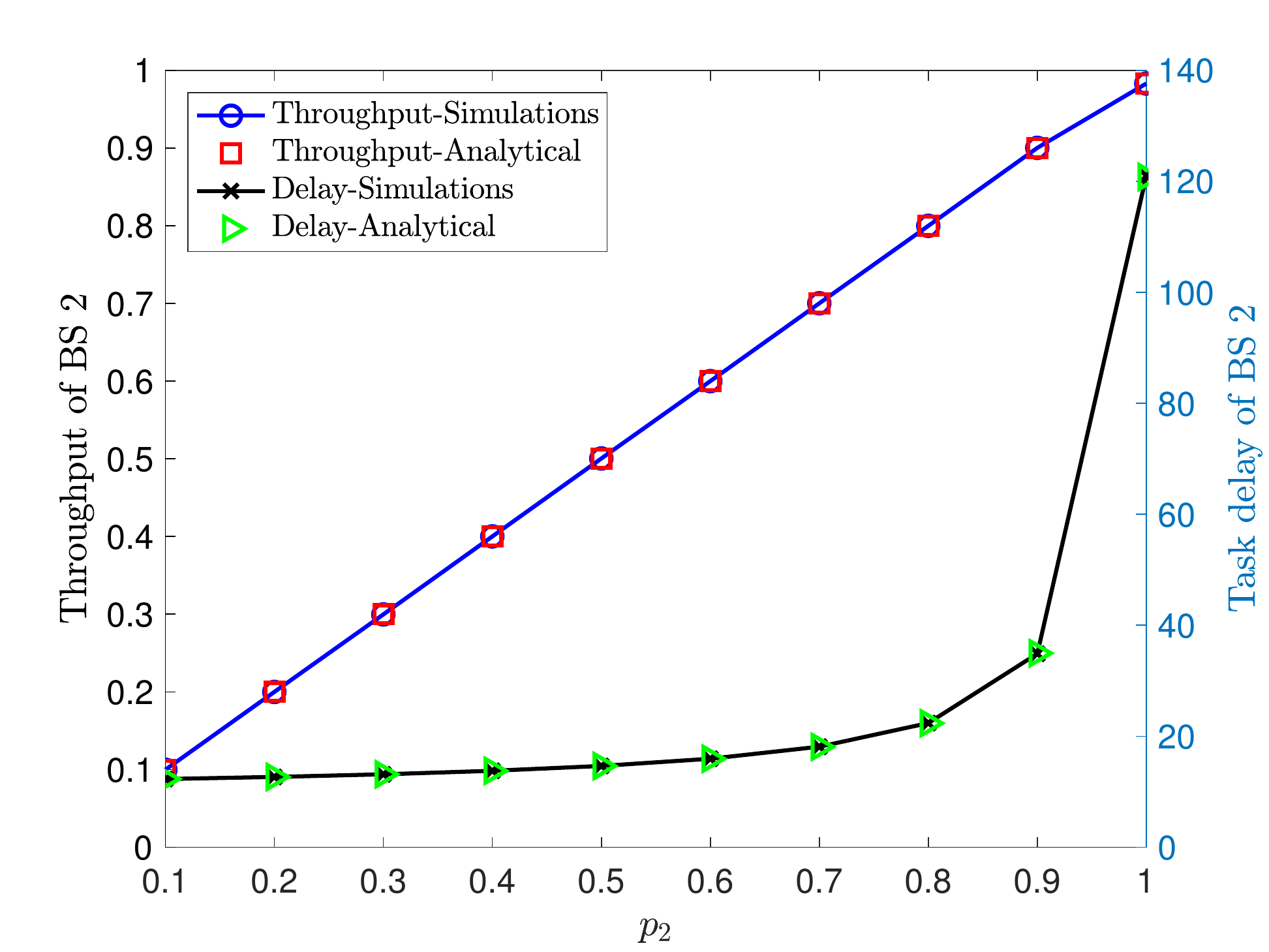}
		\label{Fig:  ScaledUpVarp2BS2}
	}
	\caption{Throughput vs delay performance. $p_{1}=0.7$. $\mu_{i}=0.5$ for $1\leq i \leq 5$ and  $7\leq i \leq 11$. $\mu_{6}=1$, $M_{i}=50$ for $1\leq i \leq 5$ and $ 7\leq i \leq 11$. $M_{6}=100$.}
	\label{Fig: ScaledUpVarp2}
\end{figure}

In this subsection, we study the performance of a scaled-up system with two base stations, BS1 and BS2. In addition, we evaluate the performance of the proposed approximation model by comparing the analytical and simulations results. 
The routing decisions for all the cases are equal to $0.5$.

In Fig. \ref{Fig: ScaledUpEqualChanges}, the results show the trade-off between  system throughput and system delay as the traffic of both base stations increases. It is easy to see that the analytical results are close to the simulation ones which validates the accuracy of the proposed model. In Fig. \ref{Fig:  ScaledUpVarp1}, we provide results that show the performance of each base station for different traffic volumes of BS1. In particular, in Fig. \ref{Fig: ScaledUpVarp1BS1}, we observe the trade-off between the throughput and delay of BS1. We see that as the arrival rate increases, the throughput of BS1 increases. However, the throughput of BS1 is not equal to the arrival rate for large values of the arrival rate because of the packet drops. 

In Fig. \ref{Fig:  ScaledUpVarp1BS2}, we provide results that show the trade-off between throughput and delay of BS2 as the arrival rate of BS1 increases. Our goal is to show how the traffic volume of BS1 affects the performance of BS2. It is clear that when the arrival rate of BS1 is larger than the arrival rate of BS2, the performance of BS2 decreases significantly.

In Fig. \ref{Fig: ScaledUpVarp2}, we provide results that show how the traffic volume  of BS2 affects the traffic of BS1. In this case, the service rates are higher than the previous case of Fig. \ref{Fig:  ScaledUpVarp1}. We observe that under this setup, the one system affects the other only in terms of delay. The maximum throughput of BS1 (equal to the arrival rate) is achieved for all the values of $p_{2}$. However, the effect on delay is more significant.

\section{Summary}
In this work, we consider an exemplary network topology with two MEC servers, a high-end server at core network, and VNF chains embedded in the servers. We model the network and provide a theoretical study on the system performance in terms of system drop rate and average number of the tasks in the system that can be useful for more general set-ups. We provide both experimental and theoretical results in order to evaluate the performance of our approximated model and as it is shown our derived model can approximate the system with high accuracy. Numerical results show that, we are able to offer some useful insights on the design of such systems or resource allocation at each server. Furthermore, we investigate numerically the routing policy that minimizes the system drop rate for different set-ups of the system. This work can be considered as an initial, but significant, step for analyzing and optimizing end-to-end delay, and throughput of such networks. The developed analysis, can provide guidelines for delay-aware routing and resource allocation schemes in similar systems.

\section*{Acknowledgment}
The authors would like to thank Manuel Stein for
numerous fruitful discussions and his valuable suggestions.
\begin{appendices}
	\section{}
	\begin{figure}[t!]
		\centering
		\subfloat[Markov chain for the server in the core.]{%
			\includegraphics[width=1\linewidth]{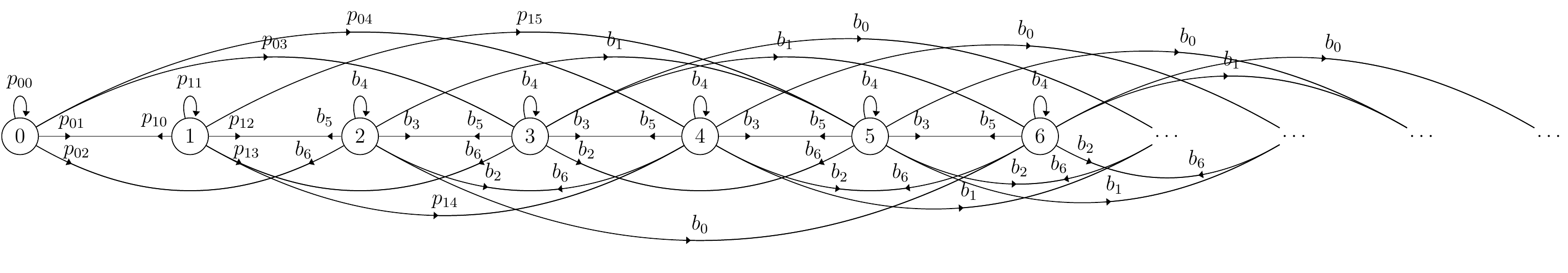}
			\label{Fig: MarkovCore}}
		\hfill
		\subfloat[Explanation of the edges.]{%
			\includegraphics[width=0.8\linewidth]{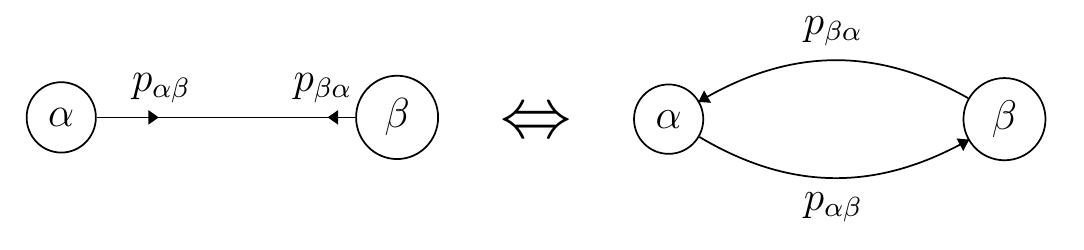}
			\label{Fig: MarkovEqu}
		}
		\caption{Stochastic model of the server in the core.}
		\label{fig:markovianicorecopy2}
	\end{figure}

We model the queue in the core server of the system  as a Markov chain. The Markov chain is depicted In Fig. \ref{Fig: MarkovEqu}. In order to facilitate the presentation, we introduce a notation which allows us to design the transitions between states in a more compact way. 
Below, we calculate the transition probabilities of the Markov chain. 
\begin{align}\nonumber
	p_{00} & =  \bar{\lambda}_{6,2} \bar{\lambda}_{6,5} \bar{\lambda}_{6,8} \bar{\lambda}_{6,11}\text{,}\\\nonumber
	p_{01} & = \lambda_{6,2} \bar{\lambda}_{6,5} \bar{\lambda}_{6,8}\bar{\lambda}_{6,11} + \bar{\lambda}_{6,2} \lambda_{6,5} \bar{\lambda}_{6,8} \bar{\lambda}_{6,11} + \bar{ \lambda}_{6,2} \bar{\lambda}_{6,5} \lambda_{6,8}\bar{\lambda}_{6,11} 
	+ \bar{ \lambda}_{6,2} \bar{\lambda}_{6,5} \bar{\lambda}_{6,8} \lambda_{6,11}\text{,} \\ \nonumber
	p_{02} & = \lambda_{6,2} \lambda_{6,5} \bar{\lambda}_{6,8} \bar{\lambda}_{6,11} + \lambda_{6,2} \bar{\lambda}_{6,5} \lambda_{6,8} \bar{\lambda}_{6,11} + \lambda_{6,2} \bar{\lambda}_{6,5} \bar{\lambda}_{6,8} \lambda_{6,11} + \bar{\lambda}_{6,2}\lambda_{6,5} \lambda_{6,8} \bar{\lambda}_{6,11}\\\nonumber
	& + \bar{\lambda}_{6,2} \lambda_{6,5} \bar{\lambda}_{6,8} \lambda_{6,11} + \bar{\lambda}_{6,2} \bar{\lambda}_{6,5} \lambda_{6,8} \lambda_{6,11}\text{,}\\\nonumber
	p_{03} & = \lambda_{6,2}\lambda_{6,5} \lambda_{6,8} \bar{\lambda}_{6,11} +  \lambda_{6,2}\lambda_{6,5}\bar{\lambda}_{6,8}\lambda_{6,11}
	+   \bar{\lambda}_{6,2}\lambda_{6,5} \lambda_{6,8} \lambda_{6,11} +  \lambda_{6,2}\bar{\lambda}_{6,5} \lambda_{6,8} \lambda_{6,11}\text{,}\\\nonumber
\end{align}
\begin{align}\nonumber
	p_{04} & =  \lambda_{6,2}\lambda_{6,5} \lambda_{6,8} \lambda_{6,11}\text{, }p_{10}  = p_{00}\Pr\left\{X\geq 1\right\}\text{, }\\\nonumber
	p_{11}  & = p_{00}\Pr\left\{X=0\right\} + p_{01} \Pr\left\{X\geq 1 \right\} \text{,}\\\nonumber
	p_{12} & = p_{01}\Pr\left\{X=0\right\} + p_{02} \Pr \left\{X\geq 1\right\}\text{, }p_{13} = p_{02}\text{Pr}\left\{X=0\right\} + p_{03}\Pr\left\{X\geq 1\right\}\text{,} \\\nonumber
	p_{14} & = p_{03} \Pr \left\{X=0\right\} + p_{04} \text{Pr}\left\{X\geq 1\right\}\text{, }	p_{15}= p_{04}\text{Pr} \left\{X=0\right\}\text{, }
	b_{0}  = p_{04}\Pr \left\{X=0\right\}\text{, }\\\nonumber
	b_{1} & =  p_{03} \Pr \left\{X=0\right\} + p_{04} \Pr \left\{X=1\right\} \text{,}\\\nonumber
	b_{2} & = p_{02}\Pr \left\{X=0\right\}  + p_{03}\Pr \left\{X=1\right\} + p_{04} \Pr\left\{X=2\right\} \text{,}\\\nonumber
	b_{3} & =p_{01}\Pr \left\{X=0\right\}  + p_{02}\Pr\left\{X=1\right\}  + p_{03}\Pr\left\{X=2\right\}\text{,}\\ \nonumber
	b_{4} & = p_{00}\Pr \left\{X=0\right\} + p_{01}\Pr\left\{X=1\right\}  + p_{02}\Pr\left\{X=2\right\}\text{,}\\\nonumber
	b_{5} & = p_{00} \Pr \left\{X=1\right\} + p_{01}\Pr\left\{X=2\right\}\text{, }b_{6}= p_{00}\Pr\left\{X=2\right\}\text{.}
\end{align}
\end{appendices}

\bibliographystyle{IEEEtran}
\bibliography{MyBib1,Bib_QL}
	
\end{document}